\documentclass[3p,letterpaper]{elsarticle}

\usepackage{graphicx}
\usepackage{xcolor}
\usepackage{csquotes}
\usepackage{epstopdf}
\usepackage{ifpdf}			
\usepackage[flushleft]{threeparttable}
\usepackage[normalem]{ulem} 
\usepackage{amsmath}
\usepackage{url}
\usepackage{epstopdf} 
\usepackage{amsmath}
\usepackage{mathrsfs}
\usepackage{tabularx} 	
\usepackage{caption}
\usepackage{letltxmacro}
\usepackage{gensymb}  
\usepackage{rotating} 
\usepackage{pdflscape} 

\usepackage{lineno}

\usepackage{comment}
\usepackage{graphicx}
\setkeys{Gin}{draft=false}
\newif\ifcorrectingmode
\correctingmodetrue

\LetLtxMacro\origcite\cite

\renewcommand{\cite}[2][]{%
  \ifcorrectingmode
  \mbox{\origcite[#1]{#2}}%
  \else
  \origcite[#1]{#2}%
  \fi
}
\journal{Advances in Water Resources}

\begin{document}

\newcommand{\fixme}[1]{ { \bf \color{red}FIX ME \color{black} #1 } }
\newcommand{\fig}{./figures/}  
%
\begin{frontmatter}
\title{A Hierarchy of Models for Simulating Experimental Results from a 3D Heterogeneous Porous Medium}
\author[1,2,3]{Daniel Vogler}
\author[3,4,5]{Sassan Ostvar}
\author[3]{Rebecca Paustian}
\author[3]{Brian D. Wood\corref{cor1}}
\cortext[cor1]{Brian D. Wood, brian.wood@oregonstate.edu}
\address[1]{ETH Zurich, Geothermal Energy and Geofluids, Institute of Geophysics, Zurich, Switzerland}
\address[2]{ETH Zurich, Transport Processes and Reactions Laboratory, Institute of Process Engineering, Zurich, Switzerland}
\address[3]{Chemical, Biological, and Environmental Engineering, Oregon State University, OR, USA.}
\address[4]{Massachusetts General Hospital, Boston, MA}
\address[5]{Harvard Medical School, Boston, MA}

\begin{abstract}

In this work we examine the dispersion of conservative tracers (bromide and fluorescein) in an experimentally-constructed three-dimensional dual-porosity porous medium. 
The medium is highly heterogeneous ($\sigma_Y^2=5.7$), and consists of spherical, low-hydraulic-conductivity inclusions embedded in a high-hydraulic-conductivity matrix.  The bi-modal medium was saturated with tracers, and then flushed with tracer-free fluid while the effluent breakthrough curves were measured.
The focus for this work is to examine a hierarchy of four models  (in the absence of adjustable parameters) with decreasing complexity to assess their ability to accurately represent the measured breakthrough curves.  The most information-rich model was (1) a direct numerical simulation of the system in which the geometry, boundary and initial conditions, and medium properties were fully independently characterized experimentally with high fidelity.  The reduced models included; (2) a simplified numerical model identical to the fully-resolved direct numerical simulation (DNS) model, but using a domain that was one-tenth the size; (3) an upscaled mobile-immobile model that allowed for a time-dependent mass-transfer coefficient; and, (4) an upscaled mobile-immobile model that assumed a space-time constant mass-transfer coefficient.  The results illustrated that all four models provided accurate representations of the experimental breakthrough curves as measured by global RMS error.  The primary component of error induced in the upscaled models appeared to arise from the neglect of convection within the inclusions.  We discuss the necessity to assign value (via a utility function or other similar method) to outcomes if one is to further select from among model options.  Interestingly, these results suggested that the conventional convection-dispersion equation, when applied in a way that resolves the heterogeneities, yields models with high fidelity without requiring the imposition of a more complex non-Fickian model.
\end{abstract}

\end{frontmatter}

\section{Introduction} \label{sec:intro}
 
In natural geological systems, highly heterogeneous materials are the rule rather than the exception. One approach for representing systems with very high variations of hydraulic conductivity is to represent the field as a set of distinct regions either through hydrofacies mapping \citep{anderson1999sedimentology}, through indicator methods \citep{knudby2005relationship}, or a combination of these two approaches \citep{lee2007geologic,bianchigeological_2017}.  A particular simplification of these models is the case of bimodal (or dual-domain) media, where only two classes of materials are present  (e.g., low conductivity {\it immobile} regions embedded in high conductivity {\it mobile} regions) \citep{vangenuchten_1976,fiori_2011,golfier_2011,davit2012correspondence,knudby2005relationship,jankovic2003effective, molinari2015analysis}. Such media can serve as an idealization of a highly heterogeneous but continuous porous medium that has been segmented into \emph{high} and \emph{low} conductivity components so that the total variance of each segment is reduced. The important hydrogeologic role of such representations has been discussed recently by \citep{molz2015advection}.  
Low-conductivity (frequently referred to by the terminology \emph{immobile}) regions are often conceptualized as being spherical or ellipsoidal in  analytical \citep[][]{coats_1964,poley1988effective,haggerty_1995,rabinovich2013dynamic,fernandez2015mathematical}, numerical \citep{finkel2016travel,lee2017longitudinal,bianchigeological_2017} and experimental \citep{zinn_2004,golfier_2011} investigations.  Ellipsoidal or spherical representations of low-conductivity inclusions have been used extensively both to represent actual structures observed in the field \citep[e.g.,][]{jussel1994transport,murphy1992influence}, and as a reasonable simplification of low-conductivity regions \citep{dagan2001solute}. The relevance in this approximation has been discussed in detail in \citet{dagan2001solute}, \citet{jankovic2006modeling} (who use ellipsoidal inclusions in their representations), and in the review by \citet{frippiat_2008}.  In a recent review article on the geological representation of heterogeneity, \citet[][p. 195]{eaton2006importance} discusses such idealizations for situations where the heterogeneity is particularly large by noting
\blockquote{
Composite media approaches, in which different heterogeneous structures of contrasting hydraulic properties, such as inclusions of different shapes, have also been used to quantify flow numerically \ldots As these methods become more widely understood, and implemented in readily available modeling codes, their application will allow a
geostatistical approach to even the most heterogeneous flow systems, a significant advance.}


Early investigations of bimodal systems accounted for the influence of immobile regions on transport phenomena by representing the immobile region with a stagnant volume fraction which is coupled to the mobile region with a constant mass transfer coefficient $\alpha$ \citep{deans_1960,deans_1963,coats_1964,rao_1980}; this idea has been extended to more general multiple-region models \citep{haggerty_1995,carrera_1998}, and models that include convection and dispersion in both regions  \citep{vangenuchten_1976,ahmadi_1998,haggerty_1995,Goltz86,haggerty_2000,golfier_2007,ginn_2017}.
Reviews of much of the literature on this topic have been presented by \citet{cherblanc_2003} and \citet{fernandez2015mathematical}. 

For bimodal representation of heterogeneous materials, the spatial domain is usually envisioned as being separated into two components: (1) a connected, high conductivity medium, and (2) a disconnected low conductivity medium.  Although in some models the low conductivity medium is assumed to be immobile, in more recent models it is assumed that convective fluxes can exist in the disconnected phase.    Because mass transfer occurs between the high- and low-conductivity regions, the resulting model can represent a range of transport behaviors from conventional convection-dispersion, to transport that appears significantly {\it non-Fickian}.  The characteristic times associated with transport in each of the two regions can span a large range if the conductivity variance in the medium as a whole is large. Such differences in transport times can result in asymmetric breakthrough curves and {\it tailing} \citep{vangenuchten_1976,haggerty_2000,zinn_2003,bianchi_2011,li_2011,fiori_2011}.  Accurate and economical descriptions of tailing phenomena have been of significant interest in hydrological applications for some time.

The objectives of this paper are (1) to describe a new set of three-dimensional experiments for solute transport in a bimodally-distributed system, and (2) to assess the ability for a hierarchy of decreasingly complex models to adequately represent the breakthrough curves from these experiments.  In particular, we are interested in the use of simplified models to simultaneously reduce the complexity (our measure of the complexity is an algorithmic one described in detail below) while maintaining fidelity with the experimental observations.

We analyze the experimental results using two strictly numerical, and two \emph{upscaled} models \citep{chastanet_2008}.  Each of these models can be described briefly as follows: (1) a  fully-detailed (i.e., resolving all heterogeneities fully) direct numerical simulation (DNS) of the entire experimental domain, (2) a fully-detailed, but domain-reduced representation of the experimental system, (3) an upscaled two-region model accounting for transience in the mass transfer process, and finally (4) an upscaled model that assumes that the mass transfer process is roughly quasi-steady (so that the mass-transfer coefficient is a constant).  One important feature of this work is that the experimental system has been highly characterized, so all models of the system are in the absence of adjustable parameters.  We examine the ability of each of these models to represent the experimental breakthrough curves, and offer some assessment as to how well reduced-complexity models perform as compared to models that represent essentially perfect information (i.e., fully-resolved DNS where the geometrical details are represented explicitly, within the bounds of experimental error).

\section{Background and Previous Work}

Bimodally-distributed media have been studied experimentally by a number of researchers; in Table 1 we have summarized the available experimental data (including this work) for both 2- and 3-dimensional systems.  We have taken particular care to report only on experiments with bimodally-distributed media and where the experimental conditions were described in sufficient detail as to make the experiments interpretable.

To address the need to capture tailing associated with bimodal media, formally averaged two-region \cite{whitaker_1999,frippiat_2008,li_2011,golfier_2011} and even multi-region \cite{davit2015theoretical} transport equations have been developed. Although transport phenomena in highly heterogeneous media have been extensively investigated numerically, studies which combine the predictive capabilities of numerical models with experimental validation at the Darcy scale are still somewhat sparse.  The most extensively characterized experiments conducted to date in bimodal media are those summarized in Table \ref{table1}.  With only two exceptions (one of which is the work reported here) these experiments were effectively 2-dimensional, and many of them have log-variance of conductivities ($\sigma_Y$) that are near unity.  The experiments detailed in this paper are unique in that they are conducted in a medium with 3-dimensional heterogeneity, and the variance is more representative of what might be observed in the field ($\sigma_Y =5.71$).

To help characterize transport phenomena in bimodal porous materials, where the two regions are denoted as the $\eta$- and $\omega$-region respectively, \citet{zinn_2004} suggested the definition of three P\'eclet numbers (as modified by \citet{golfier_2007})

\begin{equation} 
\label{eq:peclet_omega_omega}
 	Pe_{\omega \omega} = \dfrac{||\langle v_{\omega} \rangle^{\omega}||}{a} \; \dfrac{a^{2}}{D_{\omega}} = \dfrac{||\langle v_{\omega} \rangle^{\omega}|| \, a}{D_{\omega}}
\end{equation}

\begin{equation} 
\label{eq:peclet_eta_omega}
  Pe_{\eta \omega} = \dfrac{||\langle v_{\eta} \rangle^{\eta}|| a}{D_{\omega}}  \; \dfrac{a}{L}
\end{equation}

\begin{equation} \label{eq:peclet_eta_eta}
  Pe_{\eta \eta} = \dfrac{||\langle v_{\eta} \rangle^{\eta}||}{L} \; \dfrac{L^{2}}{D_{\eta}} = \dfrac{||\langle v_{\eta} \rangle^{\eta}|| L}{D_{\eta}}
\end{equation}
Here, $||\langle v_{\omega} \rangle^{\omega}||$ is the magnitude of the intrinsic velocity in the $\omega$-region  $D_{\omega}$ denotes the effective diffusivity of the solute of interest (Section \ref{sec:tracer_experiments}), $a$ is the radius of the inclusion, and $L$ denotes the characteristic distance for gradients of the concentration; conventionally, this is taken as the system length or (when applicable) the solute pulse length. To help with the interpretation of Table \ref{table1}, we note the following definitions specific to media with heterogeneities segmented into two hydraulic conductivities

\begin{align}
\bar{Y} & = \varphi_\eta \ln (K_\eta)+\varphi_\omega \ln(K_\omega) \\
\sigma^2_Y& = \varphi_\eta \left[\ln (K_\eta)-\bar{Y}\right]^2+\left[\varphi_\omega \ln(K_\omega)-\bar{Y}\right]^2
\end{align}
Note that here, $\varphi_\eta$ and $\varphi_\omega$ represent the fractions of the \emph{total} volume of the domain (fluid plus solid) occupied by the high- and low-conductivity materials, respectively.

In physical terms, $Pe_{\omega \omega}$ quantifies the relative dominance of convective and diffusive fluxes in the $\omega$-region, and $Pe_{\eta \omega}$ compares the magnitude of the convective flux in the $\eta$-region to the diffusive flux in the $\omega$-region. The final P\'eclet number, $Pe_{\eta\eta}$, compares convection and diffusion in the high-permeability matrix, and it does not usually exhibit a controlling influence on the net transport (unless it were unusually large).   In Fig.\ref{fig:pecletNumbers}, we have provided an illustration that \emph{very roughly} divides the possible combinations of $Pe_{\omega\omega}$ and $Pe_{\eta \omega}$ into five different transport schemes.  
These schemes have been discussed in detail by \citet{zinn_2004},  \citet{golfier_2007}, and \citet{heidari_2014}.  Briefly, the schemes indicate (1) the relative importance of the various convective versus diffusive processes, and (2) whether the transport time scales involved require for one- or two-equation models to represent the net transport behavior.  
This is a \emph{qualitative plot} that can be useful to generally characterize regimes, but the interpretation is the most precise when the combination of P\'eclet numbers is not near regime boundaries.  For reference, the combination of P\'eclet numbers for our experiment, and those for a number of other experiments from the literature, are plotted for comparison.  Because these data have archival value, we have also provided the associated sets of parameters for these experiments in Table \ref{table1}.

\section{Experiments} \label{sec:methods}

Because of the relative scarcity of carefully-controlled experiments in highly heterogeneous materials (particularly in 3-dimensions) we performed a set of large-scale (on the order of 1~m) experiments with a high-conductivity contrast ($\sigma_Y^2=5.71$).  As is often the case, the structure of the heterogeneity in the experimental system represented a trade-off between experimental control and interpretability versus a more realistic representation of field-like structure. As described in the Introduction, however, the relevance of bimodal materials as reasonable analogues to the field has been well established in the literature, and also reflects the most common experimental option for handling highly-heterogeneous materials \citep{zinn_2003}.   In this sense, the experimental work described below represents an extension of the work of \citet{zinn_2003} to three dimensions.

\subsection{Flow Cell}
The experimental system, illustrated in Fig.~\ref{fig:experiment_setup}, consisted of a flow cell (100~cm long, 50~cm tall, 20~cm thick), constructed of anodized aluminum, packed with a dual-porosity medium. The inlet and outlet end plates of the flow cell were engineered structures (nominally 50~cm by 20~cm anodized aluminum, with ancillary material to allow the end plates to be bolted to the flow cell), machined with groves to help distribute flow and create as uniform a pressure as possible.  Six inlet ports were installed at inlet and outlet plates, again with the goal of distributing the flow and pressure as uniformly as possible.  Finally, each end plate was machined to accept a 50~cm by 20~cm by 1/8-inch piece of sintered stainless steel to further encourage flow distribution.  At both the inlet and outlet, a manifold of 1/8-inch teflon tubing was constructed to split the flow evenly among the 6 inlet ports.  The thickness of the plate itself was about 2.5~cm, with another 2.5~cm of tubing creating the inlet and outlet manifolds.
Additional details about the construction of the flow cell, including the hydraulics of the inlet and outlet regions, is available in the thesis by \citet{harrington_2010}.

\subsection{Porous materials}
The bimodal porous medium was constructed by embedding low-hydraulic-conductivity spherical inclusions (with radius $a=2.5$~cm) in a high-hydraulic-conductivity matrix. In the material that follows, the subscripts $\eta$ and $\omega$ will be used to denote the matrix and inclusion phases, respectively. 
The materials for the two regions were composed of mono-disperse solid glass spherical particles (Potters Industries Inc., Valley Forge, PA) of two different diameters. 
The high-conductivity matrix (the $\eta$-phase) was composed of spherical particles with a nominal diameter of $2.4\pm0.4$~mm (A-240 Spheriglass). 
The inclusions (forming the disconnected $\omega$-phase) were composed of spherical glass particles with a diameter of $0.068\pm0.023$~mm (2530 Spheriglass). The spherical inclusions were created by sintering the glass particles in graphite molds for 2.5 hours at 725$^\circ$~C.  A total of 203 such inclusions were constructed.  The sintering process helped to assure that each inclusion would be geometrically similar, and dramatically simplified the process of inclusion placement in the background matrix. 

The inclusions were placed spatially using a percolation process on a simple cubic lattice.  The flow cell was divided up into 800 cubic sub-domains (of 125~cm$^3$ each, or 5~cm on a side) forming three-dimensional simple-cubic lattice.  No inclusions were placed within $5~cm$ proximity of inlet and outlet, so this left a possible 720 sub-domains that could be populated.  To populate, a random number, $n_r, 0\le n_r \le 1$, was generated for each of the sub-domains; if the random number was less than or equal to the percolation threshold, $N_p=0.333$, an inclusion was placed in that sub-domain. 


\begin{landscape}
\begin{table}
\scriptsize
\begin{threeparttable}
\caption{Experimental data for Transport Experiments in bimodal media.}
\label{table1}
\begin{tabular}{lcccccccccccccccc}
 Source& 2D/3D & Case  & $a$ & $L$  & $ \bar{\varepsilon} $ & $\varphi_\omega$ & $D_{eff}$& $\langle {\bf v}_\eta\rangle^\eta$ &  $\langle {\bf v}_\omega\rangle^\omega$& $K_\eta$ & $K_\omega$ & $\kappa$ & $\sigma_Y$ & $Pe_{\omega\omega}$ & $Pe_{\eta\omega}$ &  \\
 & &  &  ($m$)& ($m$) & - & - & ($m^2/s$) & ($m/s$) & ($m/s$) & ($m/s$) &($m/s$) & - & - & - & - &  \\
 \hline\\
 (a) \citet{golfier_2011}& 2D & 12 hour & 0.0293\tnote{*} & 1.04\tnote{\textdagger\textdagger} & 0.385 & 0.167 & $2.7\times10^{-10}\tnote{\textdagger}$& $1.02\times10^{-5}$ & $1.61\times10^{-6}$ & $2.77\times10^{-1}$ & $3.70\times10^{-2}$ & 7.5 & 1.54& 31.2   &  175.0& \\
  (b) \citet{golfier_2011}& 2D & 24 hour & 0.0293\tnote{*} & 0.52\tnote{\textdagger\textdagger} & 0.385& 0.167 & $2.7\times10^{-10}$\tnote{\textdagger} & $1.02\times10^{-5}$ & $1.61\times10^{-6}$ & $2.77\times10^{-1}$ & $3.70\times10^{-2}$ & 7.5 & 1.54 & 62.4 &   175.0 & \\
 (c) \citet{zinn_2004}& 2D &
  \begin{tabular}{@{}c@{}}$\kappa = 1800$ \\ medium flow\end{tabular}
  &  0.0127 & 0.40 & 0.425 & 0.335& $2.83\times10^{-10}$ & $3.33\times10^{-5}$ & $2.10\times10^{-8}$ & $4.10\times10^{-4}$ & $2.28\times10^{-7}$ & 1800 & 4.31 & 47.4 & 0.94 &   \\
(d)  \citet{zinn_2004}& 2D &
 \begin{tabular}{@{}c@{}}$\kappa = 1800$ \\ low flow\end{tabular}
&  0.0127 & 0.40 & 0.425 & 0.335 & $2.83\times10^{-10}$& $1.67\times10^{-5}$ & $1.04\times10^{-8}$ & $4.10\times10^{-4}$ & $2.28\times10^{-7}$ & 1800 & 4.31 & 23.8 & 0.47 &  \\
(e) \citet{zinn_2004}& 2D &
\begin{tabular}{@{}c@{}}$\kappa = 300$ \\ medium flow\end{tabular}
& 0.0127 & 0.40 & 0.425 & 0.335 & $2.83\times10^{-10}$& $3.16\times10^{-5}$ & $1.19\times10^{-7}$ & $4.10\times10^{-4}$ & $1.37\times10^{-6}$ & 300 & 3.28 & 45.0 & 5.3 &  \\
(f)  \citet{zinn_2004}& 2D &
\begin{tabular}{@{}c@{}}$\kappa = 300$ \\ low flow\end{tabular}
& 0.0127 & 0.40 & 0.425 & 0.335 & $2.83\times10^{-10}$& $1.58\times10^{-5}$ & $5.94\times10^{-8}$ & $4.10\times10^{-4}$ & $1.37\times10^{-6}$ & 300 & 3.28 & 22.5 & 2.7 &  \\
(g) \citet{zinn_2004} & 2D &
\begin{tabular}{@{}c@{}}$\kappa = 6$ \\ high flow\end{tabular}
& 0.0127 & 0.40 & 0.425 & 0.335 & $2.83\times10^{-10}$& $9.83\times10^{-5}$ & $2.00\times10^{-5}$ & $4.10\times10^{-4}$ & $6.83\times10^{-5}$ & 6 & 1.03& 141.1 & 897.5 &  \\
(h) \citet{zinn_2004} & 2D &
\begin{tabular}{@{}c@{}}$\kappa = 6$ \\ medium flow\end{tabular}
& 0.0127 & 0.40 & 0.425 & 0.335 & $2.83\times10^{-10}$& $3.33\times10^{-5}$ & $6.67\times10^{-6}$ & $4.10\times10^{-4}$ & $6.83\times10^{-5}$ & 6 & 1.03 & 47.5 & 299.2 &   \\
(i)  \citet{greiner1997}& 3D & heterogeneous & 0.0071\tnote{**} & 0.282 & 0.37 & 0.05 & $2.31\times10^{-10}$\tnote{\textdagger}&   $9.81\times10^{-5}$ & $4.04\times10^{-5}$ & $1.91\times10^{-3}$ & $7.87\times10^{-4}$ & 2.4 &0.82 & 295.1 &  2449 & \\
(j)  \citet{murphy1997}& 2D & - & 0.0143 & 0.5 & 0.37 & 0.09 & $2.69\times10^{-10}$\tnote{\textdagger}&   $5.7\times10^{-6}$ & $1.14\times10^{-7}$ & $2.76\times10^{-3}$ & $5.5\times10^{-5}$ & 50 & 3.40 & 295.1 &  2449 & \\
(k)  \citet{heidari_2014}& 2D & LCF & 0.045 & 0.2 & 0.36 & 0.203 & $5.0\times10^{-10}$ &   $-$ & $-$ & $1.08\times10^{-5}$ & $4.82\times10^{-6}$ & 2.24 & 0.58 & 1548 &  331 & \\
(l)  \citet{heidari_2014}& 2D & LC0 & 0.09 & 0.2 & 0.35 & 0.203 & $5.0\times10^{-10}$ &   $-$ & $-$ & $1.08\times10^{-5}$ & $4.82\times10^{-6}$ & 2.24 & 0.58 & 3235 &  1318 & \\
(m)  \citet{heidari_2014}& 2D & MCO & 0.09 & 0.2 & 0.33 & 0.203 & $5.0\times10^{-10}$ &   $-$ & $-$ & $1.08\times10^{-5}$ & $2.55\times10^{-6}$ & 4.23 &1.03 & 2138 &  1862 & \\
(n)  \citet{heidari_2014}& 2D & HCF & 0.045 & 0.2 & 0.33 & 0.203 & $5.0\times10^{-10}$ &   $-$ & $-$ & $1.57\times10^{-5}$ & $3.04\times10^{-6}$ & 5.16 & 1.18 & 759 &  501 & \\
(o)  \citet{heidari_2014}& 2D & HCO & 0.09 & 0.2 & 0.29 & 0.203 & $5.0\times10^{-10}$ &   $-$ & $-$ & $1.57\times10^{-5}$ & $3.04\times10^{-6}$ & 5.16 & 1.18 & 1622 &  2344 & \\
(p) {\it This work}& 3D & - & 0.025 & 1.00 & 0.39 & 0.133 & $3.64\times10^{-10}$& $8.62\times10^{-6}$ &  $3.21\times10^{-8}$ & $7.77\times10^{-4}$ & $6.67\times10^{-7}$ & 1165 & 5.71 & 14.8 &  2.2 &  \\
&&\\
 \hline
\end{tabular}
  \begin{tablenotes}
   \item[*] Radius for an equivalent volume cylinder.
   \item[**] Radius for an equivalent volume sphere.
    \item[\textdagger] Diffusion value from the literature, specific for the dissolved species used in the original study.
    \item[\textdagger\textdagger] Length of the input pulse (for input pulses shorter than the system length).
  \end{tablenotes}
\end{threeparttable}
\end{table}
\end{landscape}

\begin{figure}
	\centering
	\includegraphics[width=0.9\textwidth]{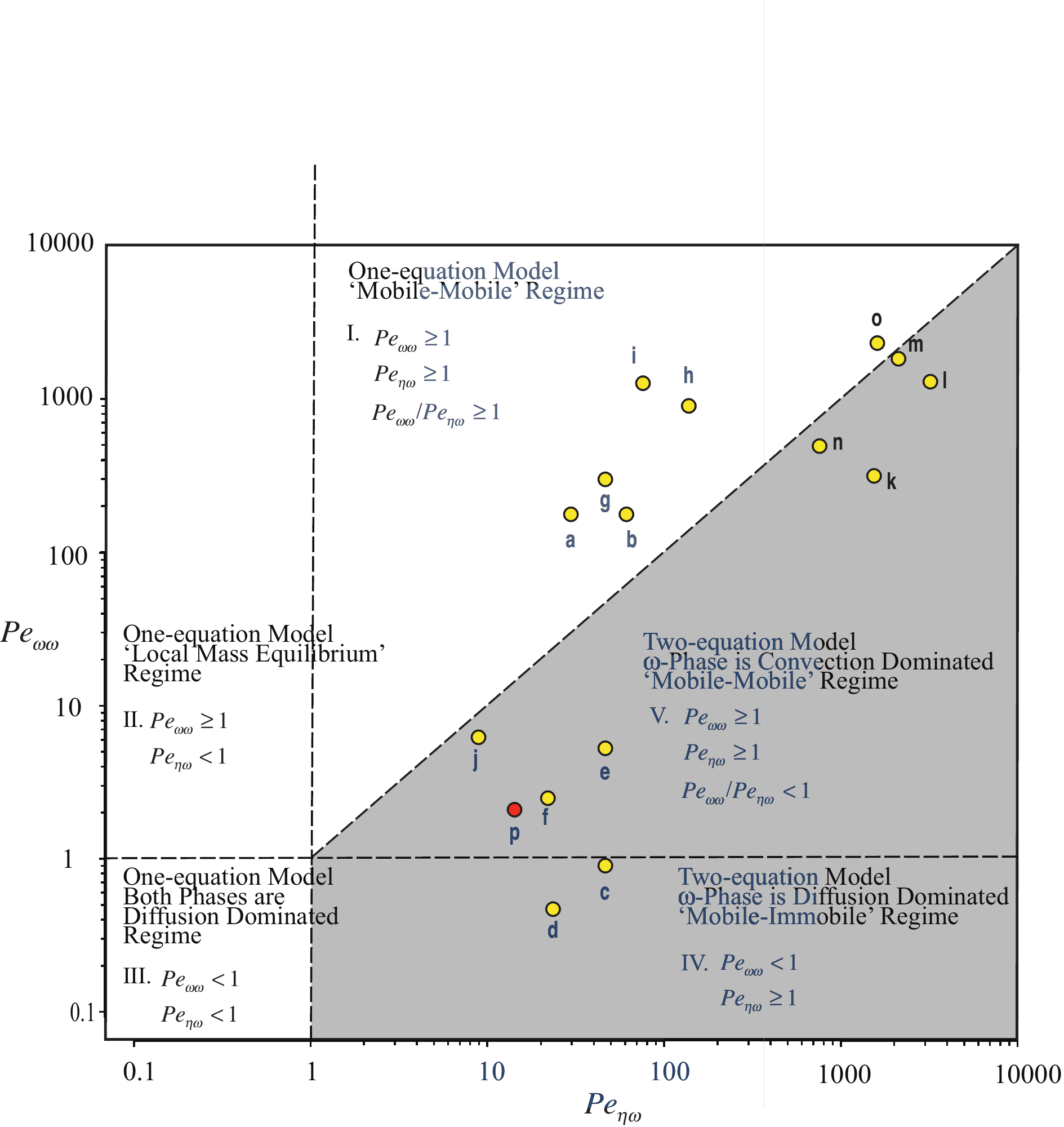}
	\caption{
	Figure of the two P\'eclet numbers $Pe_{\omega \omega}$ and $Pe_{\eta \omega}$. 
	The graph is divided into regions with respective dominant mass transport processes. 
	Experiments from other studies are included for comparison.  Letters correspond to the experimental conditions listed in Table \ref{table1}.
    }
	\label{fig:pecletNumbers}
\end{figure}

The fractions of the total volume of the domain occupied by the $\eta-$ and $\omega-$regions are denoted by $\varphi_{\eta}$ and $\varphi_{\omega}$.  Note that these two quantities are defined as including both the fluid and solid phases of the $\eta-$ and $\omega-$regions. Using the number of spheres placed and their radius, it is easy to compute that the realized fraction of inclusions was 13.3\%.  
Table \ref{tab:system_properties} provides a summary of the physical parameters that characterize the system.


The hydraulic conductivity of the coarse medium (the matrix) was measured by packing the flow cell with only the high-hydraulic-conductivity beads; the flow and pressure drop was then measured for several flow rates, and the hydraulic conductivity determined by fitting the data to Darcy's law.
The hydraulic conductivity of each of the 203 inclusions was independently measured as follows.  Individual spheres were immobilized by O-rings in a column specifically-designed for measuring the pressure drop through each sphere. 
A pressure drop was applied and the corresponding flow rate was measured to provide a measure of the \emph{relative} conductivity of each sphere.  This was then normalized by using the hydraulic conductivity of the fine medium measured in a column experiment similar to that for the coarse medium.  There is a certain amount of uncertainty in the hydraulic conductivity of the spherical inclusions because the sintering process almost certainly reduced the hydraulic conductivity of the fine porous medium compared to its non-sintered original state.  Although there was variation in the hydraulic conductivity, the average value of the hydraulic conductivity for the fine medium was used in models of the system.

\subsection{Tracer experiments} \label{sec:tracer_experiments}

Two solutions were prepared for use in experiments.  First, the tracer solution consisted of 100 ppm sodium borate (to discourage biological growth) at pH 9.3~$\pm$~0.3, to which lithium bromide ($c_0=$25 mg/L as Br$^{-}$) and fluorescein ($c_0=$1.5 mg/L) were added as inert solutes.  These two tracers were selected with the goal of using two tracers that are largely conservative, but would allow us to assure that unexpected effects (such as ion repulsion or sorption \citep{kasnavia1999fluorescent}) did not occur. Solution containing these two tracers initially saturated the flow cell. A second solution consisting of 100 ppm sodium borate was prepared for column flushing.  The column was pre-saturated with the tracer solution by pumping it through the system until concentrations measured at the inlet and outlet equilibrated.  To conduct the experiments, the tracer-free solution was injected into the inlet structures of the tracer-saturated system with an average injection rate, $Q_0$, of 45.2 mL/min \cite{harrington_2010}. 
Breakthrough concentrations of flourescein were measured at the column outlet using a model 10-AU-005-CE flow-through fluorometer (Turner Designs, Sunnyvale, CA). 
Effluent Bromide was measured by collecting effluent volume fractions (using a Gilson 223 fraction collector manufactured by Gilson, Inc., Middleton, WI); concentrations were subsequently determined using ion chromatography (DX-120 Ion Chromatograph, Dionex, Sunnyvale, CA).

As is always the case with tracer experiments in complex laboratory media, it is important to be clear about the interpretation of measured concentrations.  For our experimental system, we measured what are essentially concentrations that are flux-averaged over the exit plane of the flow cell.  Because the P\'eclet number in the coarse medium is very high ($Pe_{\eta\eta} = 23925$), we do not have to be concerned about the distinction between flux-averaged and resident concentrations as influenced by the diffusive fluxes (cf. \citet{parker1984flux}).  Thus, the concentration measured for the breakthrough curves corresponds to the following \cite{golfier_2007}

\begin{equation}
{\bar c}_{\eta} =\frac{1}{Q_0} \int_{A_{\eta,\textrm{effluent}}} {\bf n}_{\eta e}\cdot({\bf v}_\eta c_\eta)\,\, dA
\label{fluxed}
\end{equation}
In the remainder of the paper, ${\bar c}_{\eta}$ will be used to indicate either an experimentally-measured breakthrough concentration, or the appropriately-weighted (via Eq.~(\ref{fluxed})) flux averaged concentration derived from mathematical and numerical models.

\begin{figure}
  \centering
  \includegraphics[scale=0.3]{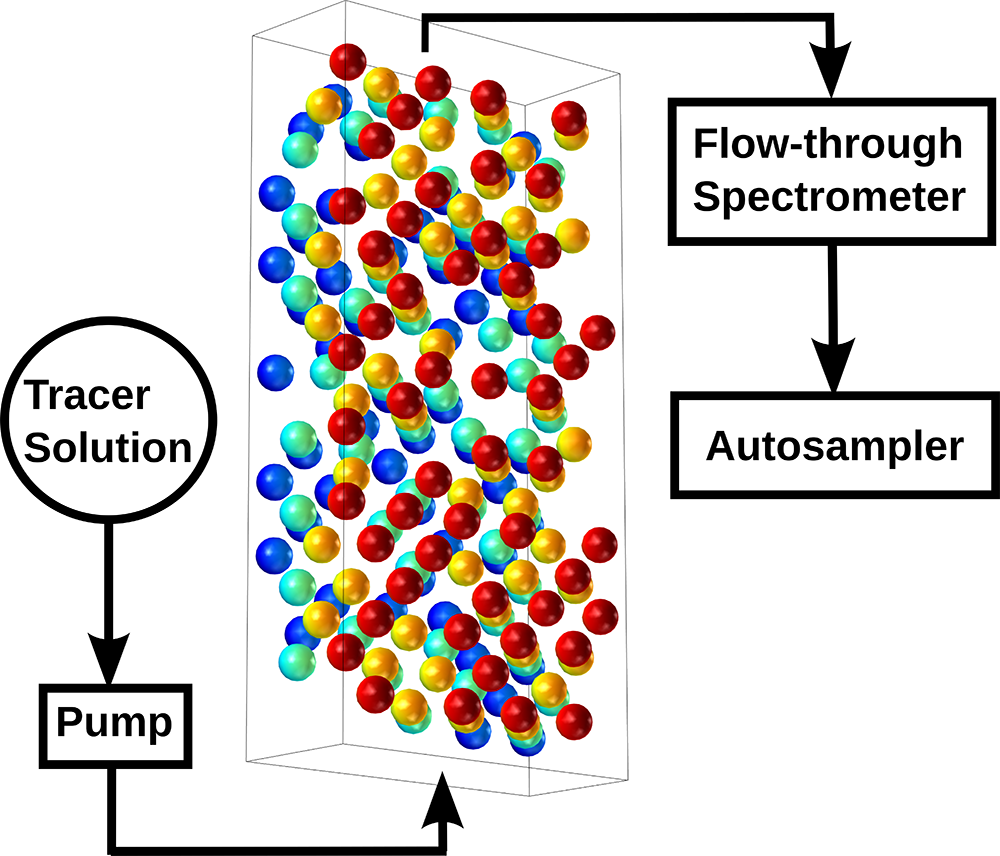}
  \caption{Schematic of experimental setup.  The flow cell was 100~cm long (in the direction of flow), 50~cm tall, and 20~cm thick.  
  Tracer solutions were pumped through the two-region medium which consisted of 203 embedded inclusions. 
  The inclusions are colored by their respective layer in the shortest system dimension (20~cm, 4 layers of inclusions). 
  Concentration is subsequently measured in the spectrometer.}
  \label{fig:experiment_setup}
\end{figure}

\begin{table}[h!]
	\caption{Material properties, transport properties, and dimensions of the experimental system. }	
	\begin{center}
    \begin{threeparttable}
		\begin{tabular}{c c c c}
			Parameter & Description & Value & Units \\ 
			\hline \\
	$Q_{0}$ & inlet flow & $45.2$ & [mL/min] \\
    $c_{0,b}$  & initial bromide concentration & 25 & [mg/L]  \\
    $c_{0,f}$  & initial fluorescein concentration & 1.5 & [mg/L]  \\
			$\varepsilon_{\eta}$ & $\eta$-region porosity & 0.4 & [$-$] \\[2pt] 
			$\varepsilon_{\omega}$ & $\omega$-region porosity & 0.31 & [$-$] \\[2pt] 
			$\varphi_{\eta}$ & volume fraction of the $\eta$-region & 0.867 & [$-$] \\[2pt] 
			$\varphi_{\omega}$ & volume fraction of the $\omega$-region & 0.133 & [$-$] \\[2pt] 
			$\rho_{\eta}$ & density of the $\eta$-region & 2.50 & [$g/cm^3$] \\[2pt] 
			$\rho_{\omega}$ & density of the $\omega$-region & 2.43 & [$g/cm^3$] \\[2pt] 
			$K_{\eta}$ & hydraulic conductivity of the $\eta$-region & 7.77$\times10^{-4}$ & [$m/s$] \\[2pt] 
			$K_{\omega}$ & hydraulic conductivity of the $\omega$-region & 0.667$\times10^{-6}$ & [$m/s$] \\[2pt] 
            $\sigma_Y^2$ & variance of the natural-log transform of the hydraulic conductivity field & 5.7 & [$-$] \\[2pt]
            $\kappa$ & ratio of high to low hydraulic conductivities & 1165 & [$-$] \\[2pt]
			$D_{m,f}$\tnote{\textdagger} & molecular diffusivity of flourescein & 4.9 $\times 10^{-10}$ & [$m^2/s$] \\[2pt] 
            $D_{m,b}$ & molecular diffusivity of bromide & 3.5 $\times 10^{-10}$ & [$m^2/s$] \\[2pt] 
			$D_{\eta}$\tnote{\textdagger\textdagger} & effective diffusivity of flourescein in the $\eta$-region & 3.77 $\times 10^{-10}$ & [$m^2/s$] \\[2pt] 
			$D_{\omega}$\tnote{\textdagger\textdagger} & effective diffusivity of flourescein in the $\omega$-region & 3.64 $\times 10^{-10}$ & [$m^2/s$] \\[2pt] 
            $\alpha_{L,\eta}$ &longitudinal dispersivity, $\eta$-region & $2.9\times10^{-2}$ & [m] \\
    $\alpha_{T,\eta}$ & transverse dispersivity, $\eta$-region  & $2.9\times10^{-3}$ & [m] \\
     $\alpha_{L,\omega}$ & longitudinal dispersivity, $\omega$-region & $1.5\times10^{-3}$ & [m] \\
    $\alpha_{T,\omega}$ & transverse dispersivity, $\omega$-region & $1.5\times10^{-4}$ & [m] \\
			$a$ & radius of spherical inclusions & 2.5 & [$cm$]\\[2pt] 
            $L$ & total flow cell length (including manifold)& 110 & [$cm$] \\[2pt]
            $\Delta L$ & inlet/outlet plate thickness (including manifold) & 5 & [$cm$] \\[2pt]
            $L_m$ & flow cell (internal) media length & 100 & [$cm$] \\[2pt] 
			$w$ & flow cell (internal) media width & 50 & [$cm$] \\[2pt] 
			$h$ & flow cell (internal) media depth & 20 & [$cm$] \\ 
			\hline 	\\
		\end{tabular}
          \begin{tablenotes}
            \item[\textdagger] From reference \citep{rani_2005}.
          	\item[\textdagger\textdagger] Computed using the conventional Maxwell relation, \citet[][Chp. 1]{whitaker_1999}.
   		\end{tablenotes}
  \end{threeparttable}
		  \label{tab:system_properties}
	\end{center}
\end{table}

\section{Hierarchy of Models for the Bimodal System}


In this section, we describe the application of a hierarchy of models, with decreasing complexity, that can be used to represent the results of the experimental system.  This data set provides an excellent opportunity to test the performance of various upscaled (information-reduced) models spanning the range from a fully-resolved direct numerical simulation (DNS) of the system, to substantially simplified models that may have significant restrictions on their range of validity.  In particular, we examine the following hierarchy of models, from the most to the least complex as follows

\begin{enumerate}
\item {\bf Model I}.  A fully-resolved and converged DNS of the entire domain, resolving each inclusion in its appropriate spatial location.

\item {\bf Model II}.  A simplified numerical model developed by conducting a fully-resolved and converged DNS, but on a domain that is substantially (one-tenth) the size of the actual flow cell domain.  The idea here is that the flow cell experiment may be much larger than a representative elementary volume (REV) of the system (in the sense described by \citet{wood_2009,wood2013volume}). The purpose of this simulation is to determine if a smaller volume of the system would still provide an REV from the perspective of the fidelity of breakthrough curves. Because there is no unique way to select and REV, and because the definition of an REV depends, in part, upon the metric chosen (e.g.,\citep{wood2015comparison}), this analysis provides some relevant details on how sensitive breakthrough curves are to the REV selection.

\item {\bf Model III}.  The upscaled model of \citet{chastanet_2008}.  This is a mass-transfer-type model that assumes only that the system is an REV, and that the relative volume fractions of the two phases (and the associated effective parameters) are known.  This particular model is somewhat more general than the mass transfer models presented by, for example, \citet{haggerty_1995}, by allowing for potential transience in the effective mass-transfer coefficient.
\item {\bf Model IV}. A simplified version of the upscaled model of \citet{chastanet_2008}.  For this model, the conditions of the system are poised such that one can assume a constant mass-transfer coefficient for the system, yielding a model that is identical to that described by \citet{haggerty_1995}.
\end{enumerate}

 Regardless of which model is selected, our analysis ultimately starts with the governing equations for the system at the microscale (for this analysis, \emph{micro scale} means the scale for which the porous material may be treated as a pre-homogenized continuum).  Because the analysis for the  fluorescein data had much higher sensitivity than the bromide data, for the remainder of the analysis the focus will be primarily on fluorescein as the species of interest.  The governing balance equations are given by

\begin{align}
  \label{eq:darcy}
  & \mbox{velocity field, $\eta-$phase}& \mathbf{v}_\eta &= -\frac{{\textbf{\sffamily \bfseries K}}_{\mathit{\eta}}}{\varepsilon_\eta } \cdot \left( \nabla p_\eta - \rho \mathbf{g}\right) \\
  & \mbox{velocity field, $\omega-$phase}& \mathbf{v}_\omega &= -\frac{{\textbf{\sffamily \bfseries K}}_{\mathit{\omega}}}{\varepsilon_\omega } \cdot \left( \nabla p_\omega - \rho \mathbf{g}\right)\\
  & \mbox{velocity B.C.1}&  {\bf n}_{\eta\omega}\cdot\mathbf{v}_\eta &= {\bf n}_{\eta\omega}\cdot\mathbf{v}_\omega \quad \,\text{at internal boundaries} \label{eq:bcv1} \displaybreak[3]\\[5pt]
  & \mbox{velocity B.C.2}&  \mathbf{v}_\eta\left.\right|_{x=0} &= \mathbf{v}_0, \quad \,\text{at flow cell inlet} \label{eq:bcv2} \displaybreak[3]\\[5pt]
  & \mbox{tracer transport,  $\eta-$phase}& \frac{\partial c_\eta}{\partial t} + \nabla\cdot (c_\eta\,\mathbf{v}_\eta) 
  &= \nabla\cdot ({\textbf{\sffamily \bfseries D}^*}_{\mathit{\eta}} \cdot \nabla c_\eta)\\[5pt]
  \label{eq:transport1}
  & \mbox{tracer transport,  $\omega-$phase}& \frac{\partial c_\omega}{\partial t} + \nabla\cdot (c_\omega\,\mathbf{v}_\omega) 
  &= \nabla\cdot ({\textbf{\sffamily \bfseries D}^*}_{\mathit{\omega}} \cdot \nabla c_\omega)\\[5pt]
  \label{eq:transport2}
  & \mbox{I.C.}& \ c_{\eta}|_{t=0} &= c_{\omega}|_{t=0} = c_{0,f} \\[5pt]
  & \mbox{interfacial B.C.}& \ c_{\omega} &= c_{\eta} \quad \,\text{at internal boundaries} \label{eq:bc} \displaybreak[3]\\[5pt]
  & \mbox{inlet B.C.}& \ c_{\eta}|_{x=0} &= 0  \quad \,\text{at flow cell inlet}\\
  & \mbox{outlet B.C.}&  \ \frac{\partial c_{\eta}}{\partial x}|_{x=L} &= 0 \quad \,\text{at flow cell outlet} \label{eq:outlet}
\end{align}
Here $\textbf{v}$, $p$, $c_{\eta}$, and $c_{\omega}$ are the Darcy-scale pointwise velocity, pressure, and the scalar concentrations in each region, respectively, and ${\bf g}$ is gravitational acceleration. 

This system of equations is valid in both regions and provides a complete description of the system at all the scales involved in the experiments. 
The corresponding hydraulic conductivity at the Darcy scale is specified by ${\textbf{\sffamily \bfseries K}}_{\eta}=\textbf{\sffamily \bfseries I}K_\eta$, ${\textbf{\sffamily \bfseries K}}_{\omega}=\textbf{\sffamily \bfseries I}K_\omega$.  For the effective dispersion tensors ${\textbf{\sffamily \bfseries D}^*}_{\eta}$ and ${\textbf{\sffamily \bfseries D}^*}_{\omega}$, the following relations hold 

\begin{align}
{\textbf{\sffamily \bfseries D}^*}_{\eta} &= \textbf{\sffamily \bfseries I}D_\eta
+\alpha_{L,\eta} ||\langle {\bf v}_{\eta} \rangle^{\eta}||
\left[ {\begin{array}{*{20}{c}}
  1&0&0 \\ 
  0&0&0 \\ 
  0&0&0 
\end{array}} \right]
+
\alpha_{T,\eta}||\langle {\bf v}_{\eta} \rangle^{\eta}||
\left[ {\begin{array}{*{20}{c}}
  0&0&0 \\ 
  0&1&0 \\ 
  0&0&1 
\end{array}} \right] \\
{\textbf{\sffamily \bfseries D}^*}_{\omega} &= \textbf{\sffamily \bfseries I}D_\omega
+\alpha_{L,\eta}||\langle {\bf v}_{\omega} \rangle^{\eta}||
\left[ {\begin{array}{*{20}{c}}
  1&0&0 \\ 
  0&0&0 \\ 
  0&0&0 
\end{array}} \right]
+
\alpha_{T,\omega} ||\langle {\bf v}_{\omega} \rangle^{\omega}||
\left[ {\begin{array}{*{20}{c}}
  0&0&0 \\ 
  0&1&0 \\ 
  0&0&1 
\end{array}} \right]
\end{align}
Note that the materials are assumed to be structurally  isotropic, so the tensors take particularly simple forms.  The inlet and outlet plates on the flow cell extended the effective length of the domain.  Although the actual amount of additional dispersion added by these flow structures was not measured, we made a modeling choice to represent them as adding an additional 5~cm to the length at both the inlet and the outlet; this is consistent with the physical size of the inlet and outlet structures.  The dispersivity associated with the inlet and outlet plates was assumed to be the same as the coarse medium.  A full listing of the flow cell dimensions and properties are provided in Table~\ref{tab:system_properties}.\\

\subsection{Numerical Models}
Two different direct numerical simulations of the experimental system were constructed.  The first employed a full representation of the complete geometry as it was set up in the experiments (the hexahedral volume of the flow cell plus each of the 203 inclusions-- 20 cm $\times$ 50 cm $\times$ 100 cm ). The second numerical model was constructed as a similarly resolved model, but with a substantially-reduced domain size (10 cm $\times$ 10 cm $\times$ 100 cm). 

The finite elements package COMSOL Multiphysics 5.3 was used to numerically solve Eqs.~(\ref{eq:darcy})-(\ref{eq:outlet}) on this geometry.  An illustration of the geometry is given in Fig$.$~\ref{fig:experiment_setup}.  
Direct numerical simulations of the system were conducted by meshing volumes (the coarse matrix and inclusions) using a tetrahedral mesh; the matrix and inclusions were separated by a boundary-fitted mesh. The interfacial boundary condition indicated by Eq.~(\ref{eq:bc}) was applied at this interfacial boundary.

Convergence of the numerical model was determined by conducting a conventional grid convergence analysis.  In short, a sequence of simulations with decreasing mesh size were computed, and a global error metric was computed for each increasingly resolved mesh. The simulations were considered converged when the error metric was below a pre-determined target value.  Details of the convergence analysis are provided in the Appendix.

\begin{figure}
  \centering
  \includegraphics[scale=0.5]{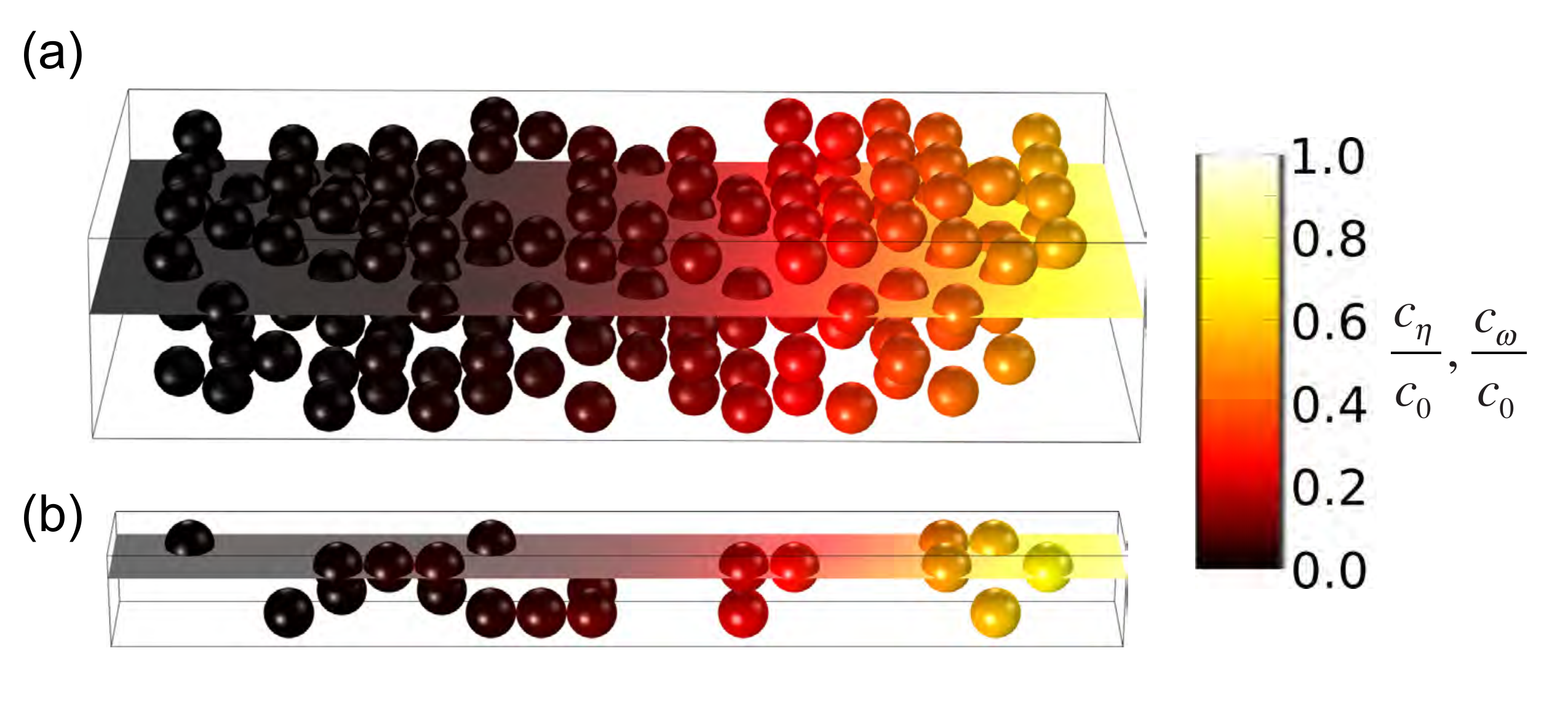}
  \caption{(a) The domain associated with the complete representation of the geometry for the flow cell DNS computations.  (b) The domain associated with the reduced domain representation of the flow cell. The color scale illustrates the normalized concentrations for $t=10$ hours.  Flow is from left to right.
   }
  \label{fig:flowcells}
\end{figure}

{\textbf{Model I. Fully-resolved direct numerical simulation of the entire flow cell.}}  The most resolved model that can be adopted for this particular system on the Darcy scale is a fully-converged numerical simulation that solves Eqs.~\ref{eq:darcy}-\ref{eq:outlet} over a domain that represents the entire experimental system (to within our abilities to measure the system parameters).  For reference, the domain for the complete representation of the geometry of the experimental system is illustrated in Fig.~\ref{fig:flowcells}a.  The advantage of such a model is that, assuming that  Eqs.~\ref{eq:darcy}-\ref{eq:outlet} are correct, there is no \emph{upscaling} involved in the representation of the system.  The primary disadvantage, however, is that such a model is maximal in terms of the amount of degrees of freedom required to represent the system.  Although this is possible for an experimental system such as this, it unlikely to be possible for actual porous materials encountered in the environment (or, even, in engineered applications).  Thus, the fully-resolved DNS is not a parsimonious model, and it may provide more information than is technically required.   As an example, the DNS model presumably provides an accurate time history of the concentration at each point in the flow cell.  This kind of information is of interest if the pointwise concentration in the medium were of relevance.  However, in many cases (including these experiments) the breakthrough curve is the only measured data that is available.  Hence, although one may predict good estimates of the concentration for each point in the domain, these are of little value if the goal of the model is to predict the breakthrough curve.  Therefore, the fully-resolved DNS model is an accurate model, but one that generates ancillary data that is not of direct relevance to the goal of computing an accurate breakthrough curve.

{\textbf{Model II. Fully-resolved direct numerical simulation of a reduced-domain.}  In model I, a fully-resolved representation of the full experimental system was developed.  For model II, we examined options for replacing the actual domain by a reduced-size realization of the domain.  The idea for this model was to roughly determine what might constitute a representative volume (REV) for the simulation of breakthrough curves.  As a guiding principle, we assumed that the volume should at least have the correct spatial statistics, which would define an REV with some quantifiable error \cite{wood2013volume}). In practice, however, uniquely determining an REV is a difficult task, and it depends in part upon the processes modeled and the required fidelity of the results. Our approach was to make a sequence of reasonable assumptions regarding the required size; these were based, in part, on previous experience with such systems.  Thus, our process was to (1) make reasonable assumptions, and then (2) check to assure that the reduced model was an REV (although it may not be the smallest such volume) as measured by fidelity with the breakthrough curves.  

For this model, we chose a domain with the full domain length (length = 100~cm), but where the width and height of the domain were set to 10~cm each. This reduction decreases the volume of domain by a factor of 10 in comparison to the fully-resolved domain. The width and height of the reduced-domain were chosen such that they were at least the size of two times the inclusion diameter.  This choice was made as a compromise between increasing fidelity for the derived effective parameters, and reducing the size of the simulated domain.  Although further size decreases in the transverse directions could be explored, increased uncertainty in the value of the effective parameters would result.  Our results indicate that our particular choice for reduction was, in fact, a reasonable compromise.

The volumetric ratio of inclusions to the total domain size was kept the same as in the experimental setup and the fully-resolved domain ($V_{\omega}/V_{total}=0.133$). Inclusions were randomly placed throughout the reduced-domain with a script generated with MATLAB \cite{vogler_2012}. As in the experiment and the fully-resolved model, no inclusions were placed within 5~cm of the inlet or outlet. For reference, the domain for the complete representation of the geometry of the experimental system is illustrated in Fig.~\ref{fig:flowcells}b.  Our reduced domain size was selected on the basis of previous experience \cite{golfier_2007}; no effort was made to determine if this was a minimal REV. Note that for both models I and II, we computed the value of ${\bar c}_{\eta}$ at the effluent plane.  This computed concentration is in principle equivalent to that measured experimentally.\\

\subsection{Upscaled Models} \label{sec:uspcaling}

We considered two additional reduced models for describing the bimodal system behavior.  Each of these models has been spatially averaged to develop upscaled balance equations.
Spatially-averaged models of mass transfer in two-region media have previously been developed in several studies \citep{whitaker_1999,quintard_1993a,wood_2003,cherblanc_2003,chastanet_2008}. 
We apply the development by \citet{chastanet_2008}; interested readers are referred to that work for details of the upscaling process. \citet{chastanet_2008} develop three upscaled models where the functional form of the effective mass transfer terms represent different levels of coupling between micro- and macroscales.  The models in decreasing order of complexity are: (1) a fully-coupled model with a non-local-in-time mass transfer term, (2) a decoupled transient model with linear mass transfer that depends upon a time-varying effective mass transfer coefficient, and (3) a quasi-steady model with a constant (asymptotic) mass transfer coefficient.  The uncoupled and the quasi-steady models (III and IV below) are both investigated in the present study; we discuss the fully-coupled model for completeness.  Note that in all of these models, it is assumed that convection in the inclusions can be neglected relative to diffusion.  This is not due to a limitation of the averaging method; rather, this approximation is required for one to develop an explicit series solution for the effective mass transfer coefficient.  Upscaled models that include convection in the inclusions are available \cite{ahmadi_1998}, but they require much more computation to develop solutions.  Ultimately, the approach of \citet{chastanet_2008} was selected as a modeling choice that appeared to be consistent with the physical system being analyzed while also providing substantial reduction in model complexity.

The fully coupled model of \citet{chastanet_2008} is stated as

\begin{align}
\label{eq:coupled_macro_eta}
\mbox{$\eta$-phase:\hspace{30mm}} \nonumber \\[10pt]
\varepsilon_{\eta} \varphi_{\eta} \frac{\partial \langle c_{\eta}\rangle^{\eta}}
{\partial t} &= \nabla \cdot \left( \varepsilon_{\eta} \varphi_{\eta}
{\textbf{\sffamily \bfseries D}}_{\eta\eta}^{**} \cdot \nabla \langle c_{\eta} \rangle^{\eta} \right)
 - \varepsilon_{\eta}
\varphi_{\eta} \langle \mathbf{v}_{\eta} \rangle^{\eta} \cdot \nabla \langle
c_{\eta} \rangle^{\eta} - W(t)  \\[15 pt]
\mbox{\hspace{-10mm} $\omega$-phase:\hspace{30mm}} \nonumber \\[10pt]
\varepsilon_{\omega} \varphi_{\omega} \frac{\partial \langle
c_{\omega}\rangle^{\omega}} {\partial t} &= \nabla \cdot ( \varepsilon_{\omega}
\varphi_{\omega} {\textbf{\sffamily \bfseries D}}_{\omega}^{} \cdot \nabla \langle c_{\omega}
\rangle^{\omega} ) + W(t)
\label{eqomega}
\end{align}
where

\begin{equation}
  \label{eq:bigW}
  W(t) = \varepsilon_{\eta} \varphi_{\omega} \; \frac{d}{dt} \left ( \int_{0}^{t} B(t - \tau) \; C_{\eta \omega}(\tau) \; d\tau \right)
\end{equation}

\begin{equation}
  \label{eq:bigC}
  C_{\eta\omega} = \left ( \left<c_{\eta}\right>^{\eta} - \left< c_{\omega} \right>^{\omega} \right)
\end{equation}
Here, the intrinsic averaged macroscale concentrations for the $\eta$- and $\omega$-regions are given by $\left<c_{\eta}\right>^{\eta}$ and $\left< c_{\omega} \right>^{\omega} $, respectively; the intrinsic averaged macroscale velocity is given by $\langle \mathbf{v}_{\eta} \rangle^{\eta}$; ${\textbf{\sffamily \bfseries D}}_{\eta\eta}^{**}$ is the effective total dispersion tensor for the $\eta$-phase;  ${\textbf{\sffamily \bfseries D}}_{\omega}$ is the effective diffusion tensor for the $\omega$-phase (${\textbf{\sffamily \bfseries D}}_\omega= {\textbf{\sffamily \bfseries I}} D_\omega $); and $W(t)$ is the effective mass transfer function \cite{chastanet_2008}. The quantity $B(t)$ is a kernel function depending on the geometry, initial conditions, and physical parameters describing the system. For clarification, we note that for the case examined here (zero convection in the $\omega$-phase),  we have the condition that ${\textbf{\sffamily \bfseries D}^*}_\omega={\textbf{\sffamily \bfseries D}}_\omega= {\textbf{\sffamily \bfseries I}} D_\omega$; thus, we have used the symbol ${\textbf{\sffamily \bfseries D}}_\omega$ in Eq.~\ref{eqomega}.

For this model, the functional form of the mass transfer source/sink term depends on the convolution of a kernel function $B$, which carries information about the physics and geometry of the problem at the Darcy scale (Eq~(\ref{eq:bigW})), with the macroscale mass transfer driving force (Eq~(\ref{eq:bigC})).  This general form is technically necessary only for conditions where the time scale for solute transport out of the inclusions is on the order of the one for transport through the coarse material.  Under many reasonable sets of conditions, this general model can be somewhat simplified while still maintaining accuracy.  In the following, we describe two of these models.  These two simplified representations were used to model the breakthrough curve of the effluent concentration from the flow cell.  Note that because of the large P\'eclet number associated with the $\eta$-phase, for this mathematical model under the experimental conditions there is essentially no difference between the concentrations $\left<c_{\eta}\right>^{\eta}$ and ${\bar c}_{\eta}$.  For convenience in presentation, we will use the symbol ${\bar c}_{\eta}$ when comparing experimental concentrations with those developed from Models III and IV with the understanding that they are equivalent.

{\textbf{Model III. Two-equation, time-local model (decoupled model).}} 
The first simplification arises under conditions where the timescales of variations of $C_{\eta\omega}$ are large compared to the time scale associated with the kernel $B$. 
When this is true the averaged concentration term, $C_{\eta \omega}$, can be carried outside the integral. 
Under these conditions, the macroscale concentrations are then {\it decoupled} from the Darcy-scale physics (represented by the kernel $B$), and the expression in Eq.~(\ref{eq:bigW}) simplified to the familiar linear form with a time-varying mass transfer coefficient

\begin{equation}
  \label{eq:linear_alpha}
  W(t) \approx  \underbrace{\varepsilon_{\eta} \varphi_{\omega} B(t)}_{\alpha(t)} C_{\eta\omega} 
\end{equation} 
Equation~(\ref{eq:linear_alpha}) transforms the macroscale balance laws into the more familiar form

\begin{align}
\label{eq:linear_macro_eta}
\mbox{$\eta$-phase:\hspace{40mm}} \nonumber \\[10pt]
 \varepsilon_{\eta} \varphi_{\eta} \frac{\partial \langle c_{\eta}
\rangle^{\eta}} {\partial t}
 = &\nabla \cdot \left( \varepsilon_{\eta} \varphi_{\eta}
{\textbf{\sffamily \bfseries D}}_{\eta\eta}^{**} \cdot \nabla \langle c_{\eta} \rangle^{\eta} \right)
- \varepsilon_{\eta} \varphi_{\eta} \langle \mathbf{v}_{\eta} \rangle^{\eta}
\cdot \nabla \langle c_{\eta} \rangle^{\eta}\\
&- \alpha(t)\left( \langle c_{\eta} \rangle^{\eta} - \langle c_{\omega}
\rangle^{\omega} \right) \notag \\[20pt]
 \mbox{$\omega$-phase:\hspace{40mm}} \nonumber \\[10pt]
 \varepsilon_{\omega} \varphi_{\omega} \frac{\partial \langle c_{\omega}
\rangle^{\omega}} {\partial t} 
 = &\nabla \cdot \left( \varepsilon_{\omega} \varphi_{\omega}
{\textbf{\sffamily \bfseries D}}_{\omega}^{} \cdot \nabla \langle c_{\omega} \rangle^{\omega} \right)
+ \alpha(t)\left( \langle c_{\eta} \rangle^{\eta} - \langle c_{\omega}
\rangle^{\omega} \right)
\end{align}

\begin{figure}
	\centering
	\includegraphics[width=0.5\textwidth]{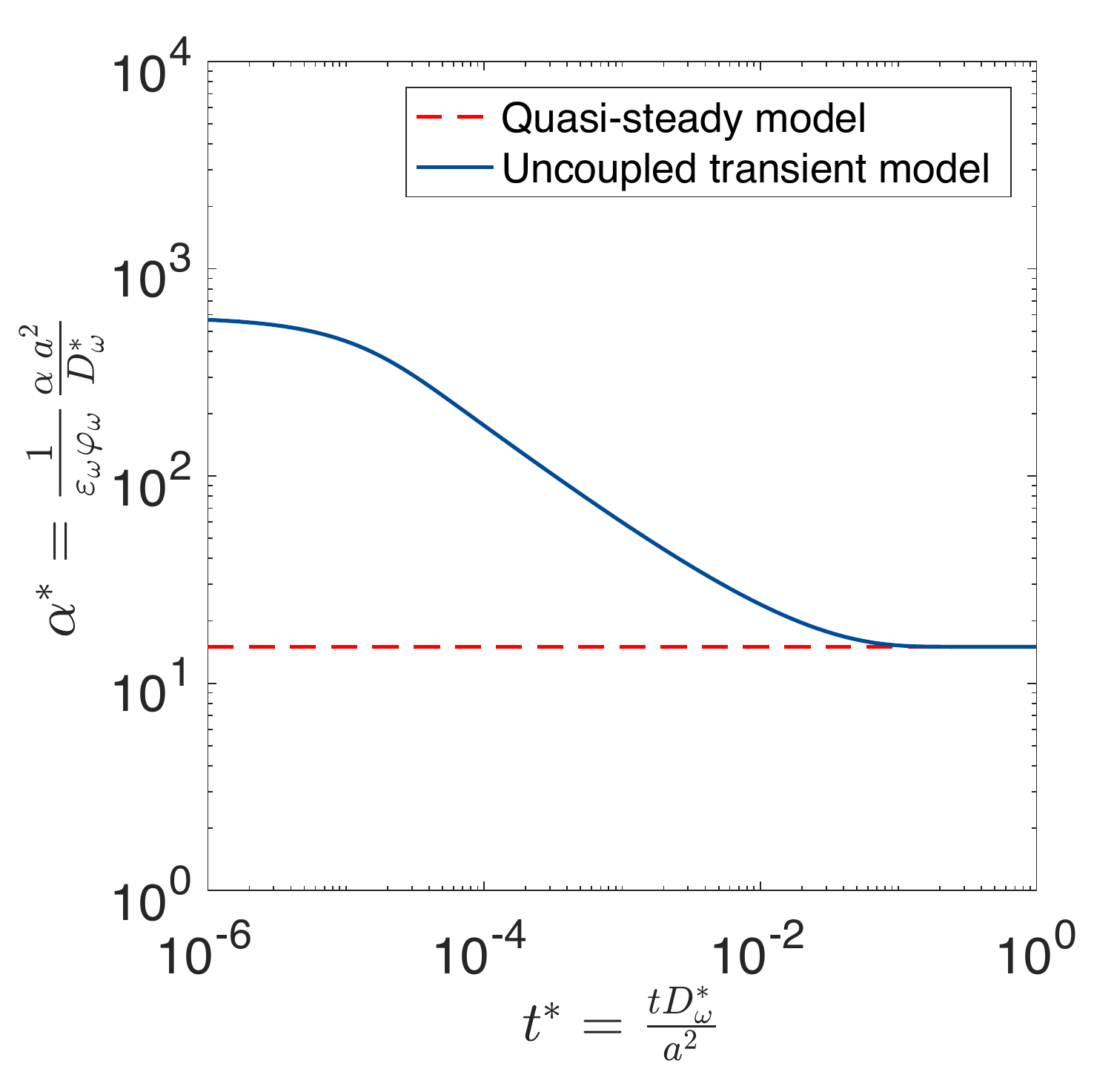}
	\caption{
	Mass transfer coefficient as computed using the uncoupled (Eq$.$~(\ref{eq:alpha_uncoupled})) and quasi-steady (Eq$.$~(\ref{eq:alpha_qs})) models.
	}
	\label{fig:alpha}
\end{figure}

\noindent\citeauthor{chastanet_2008} \cite{chastanet_2008} provide a closed-form expression for $\alpha$ by solving the closure problems associated with Eq$.$~(\ref{eq:linear_macro_eta}) over a representative elementary volume (REV) using Fourier transforms. 
The time-dependent solution is

\begin{equation} 
  \label{eq:alpha_uncoupled}
  \alpha (t) = 15\frac{D_\omega \varepsilon_{\omega} \varphi_{\omega}} {a^{2}} + 6\frac{D_{\omega} \varepsilon_{\omega} \varphi_{\omega}} {a^2} \sum_{n=1}^{\infty} \exp{\left( -\frac{q_n^{2} D_{\omega}} {a^2} t \right)}
\end{equation}
where $q_{n}$ are the non-zero positive roots of

\begin{equation} \label{eq:qn}
  \tan{q_n} = \frac{3q_n} {3 - q_n^2}
\end{equation}
For reference, a normalized plot of the transient value of $\alpha(t)$ appears in Fig.~\ref{fig:alpha}.

{\textbf{Model IV. Two-equation, quasi-steady model.}} This model is a simplification of Model III where the time asymptotic solution (the leading term in Eq$.$~(\ref{eq:alpha_uncoupled})) is considered.  For these conditions, the mass transfer coefficient is given by the constant

\begin{equation}
	\alpha(t) \approx 15\frac{D_\omega \varepsilon_{\omega} \varphi_{\omega}} {a^{2}}
	\label{eq:alpha_qs}
\end{equation}
The value of this constant is also plotted in Fig.~\ref{fig:alpha}.

Solutions for the breakthrough curves were generated using Models III and IV by treating each as a coupled (mobile-immobile) system of 1-dimensional advection-dispersion equations. The values for model parameters are listed in Table \ref{tab:system_properties}.  The transport properties reported in Table \ref{tab:system_properties} were determined a priori and independent of the experimental results.  The hydraulic conductivity of the low-conductivity region was obtained from measurements with the material used to construct the inclusion, which was tightly packed ($K_{\omega} = 6.67 \times 10^{-7}~m/s$) \cite{harrington_2010,vogler_2012}. 

For both models III and IV, the effective dispersion tensor, ${\textbf{\sffamily \bfseries D}}_{\eta \eta}^{**}$, is an \emph{upscaled}  parameter representing the dispersion in the coarse material as it is influenced by the presence of the inclusions.  We determined the effective dispersion tensor by numerically computing the solution to a closure problem posed for a single (simple) unit cell following \citet[][Chp. 3]{whitaker_1999};  

\begin{equation}\label{eq:deff}
{\textbf{\sffamily \bfseries D}}_{\eta \eta}^{**} = D_{\eta} \left({\textbf{\sffamily \bfseries I}} + \dfrac{1}{V_{\eta}}  \int_{A_{\eta\omega}} \mathbf{n}_{\eta\omega} \otimes \textbf{b}_{\eta} dA \right) - \langle \tilde{\mathbf{v}}_{\eta} \otimes\mathbf{b}_{\eta} \rangle^{\eta}
\end{equation}
Here, the ancillary vector field ${\bf b}_\eta$ is related to the deviation concentration $\tilde{c}_\eta = {\bf b}_\eta \cdot \langle c_\eta \rangle^\eta$; additional information about the particular balance equations met by ${\bf b}_\eta$ are available in \citet[][Chp. 3]{whitaker_1999}.  Our particular computation had explicitly made the approximation that the $\omega$-phase may be treated as being impermeable for the purposes of determining the effective dispersion tensor for the $\eta$-phase. 
The geometry of the unit cell (Fig$.$~\ref{fig:closure}) consisted of a sphere centered in a cube, with the volume of the sphere reflecting the total volume of the inclusions in the experimental system (Table \ref{tab:system_properties}, $\varphi_{\omega} = 0.133$). 
Subsequently, the closure variable $\mathbf{b}_{\eta}$ was calculated using COMSOL Multiphysics\textsuperscript{\textregistered}; the resulting values for the dispersion tensor components, the dispersivities, and other derived parameters are reported in Table~\ref{table:velocities}. 
A detailed derivation of $\mathbf{b}_{\eta}$ can be found in \citet{ahmadi_1998}.  This approach does require that one know something about the \emph{geometry} of the low-conductivity heterogeneities, and that this geometry be considered to be representative.  However, this approach is not limited to spherical or ellipsoidal inclusions; essentially any geometry for low-conductivity heterogeneities can be accounted for in the closure.  The primary restriction is that the low-conductivity heterogeneities be approximately spatially stationary in distribution (i.e., that some notion of a representative structure can be established).

\begin{figure}
\centering
	\includegraphics[width=0.8\textwidth]{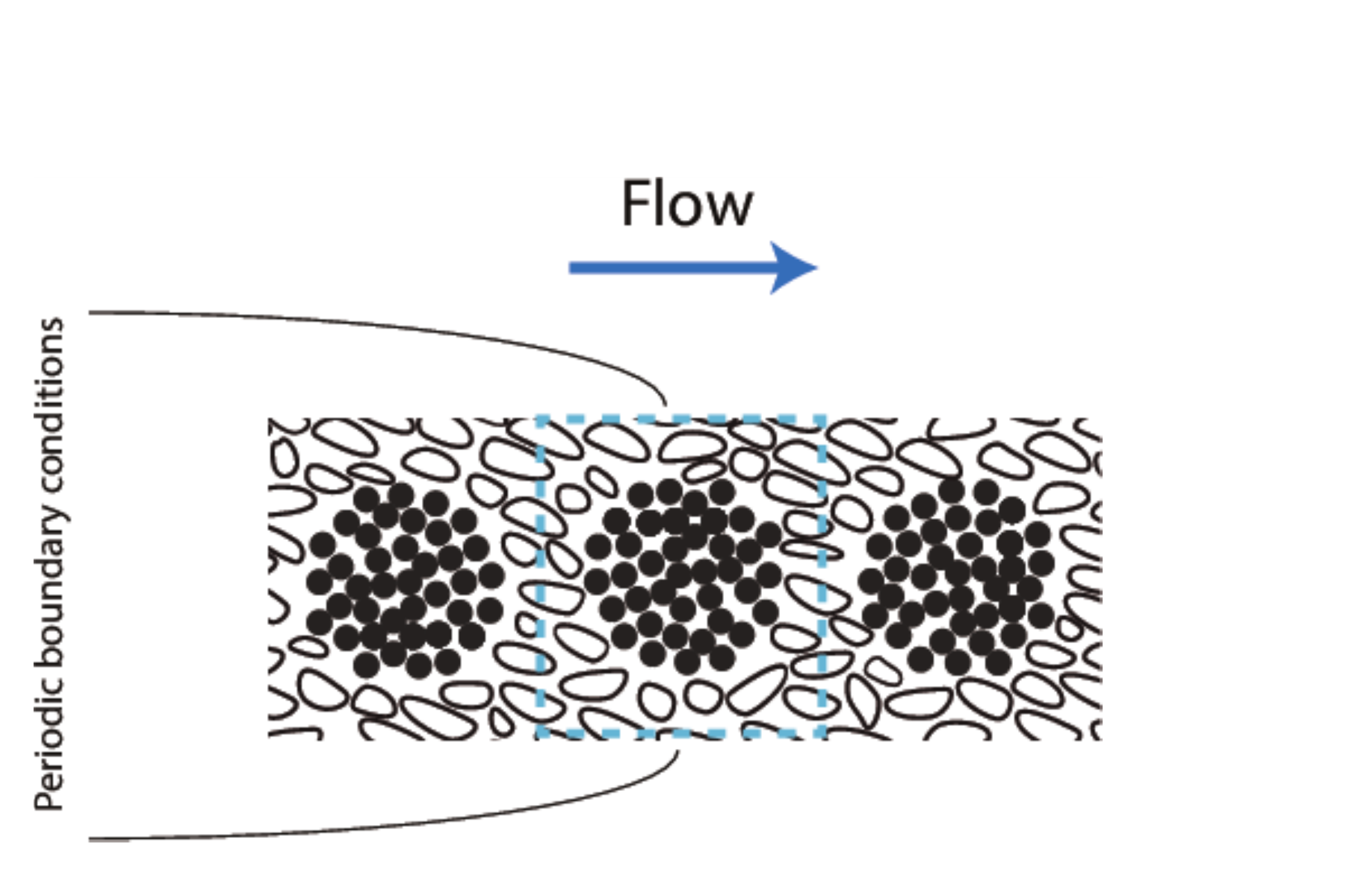}
	\caption{Geometry used in the closure problem to compute the effective dispersivity tensor. A 2D representation is depicted for simplicity, while the closure problem was computed on a 3D geometry. 
	}
	\label{fig:closure}
\end{figure}

\section{Results} \label{sec:results}


Early- and late-time experimental breakthrough curves for bromide and flourescein are illustrated in Fig$.$ \ref{fig:btc_model_comparison}. 
Tracer concentrations are normalized with respect to the initial (saturated) concentration, $c_{0,f}$ or $c_{0,b}$, and time is non-dimensionalized using the characteristic timescale of diffusion in the inclusions ($\tau_{D} = a^2 / D^{*}_{\omega}$). 
Here, \emph{early time} refers to the period of time where the bulk of the solute is being flushed out of the matrix ($0 < t^* < 0.04$); \emph{late time} refers to the period where solute is being transported primarily out of the inclusions and the outlet concentration is decreasing asymptotically ($0.04 < t^* < 0.4$). 
The observed breakthrough curve behavior thereby resembles previous studies on high-conductivity contrast domains \citep{haggerty_1995,zinn_2004}, where a sharp, initial decline in concentration after the flushing of the high-conductivity domain is followed by slower concentration decay while the solute leaves the low-conductivity inclusions. 
As Fig$.$ \ref{fig:btc_model_comparison}(b) shows, the experiment using flourescein provided more data for late-time behavior; due to the sensitivity of the method, concentrations up to $\sim$3 orders of magnitude smaller than the inlet concentration could be measured. For bromide measurements, the accuracy of the measurement technique allowed measurements only to within $\sim$2 orders of magnitude smaller than the inlet concentration. 
Figure~\ref{fig:btc_model_comparison} also illustrates that the two tracers follow very similar trends but show several small deviations from one another at both early and late times.  However, both early- and late-time deviations of bromide from fluorescein are within a 95\% confidence interval for bromide; thus, they are not statistically significant.  The late-time deviations are most likely additionally influenced by approaching the noise floor for measurements using the HPLC.  
The spatial distribution of the simulated tracer concentrations in the system are illustrated in Fig.~ \ref{fig:conc_field_visualization}.  This plot is useful to get a sense for the disparity in time scales for transport in the matrix versus the inclusions. 
      

The magnitude of the intrinsic velocities $||\langle v_{\omega} \rangle^{\omega}||$ and $||\langle v_{\eta} \rangle^{\eta}||$ were computed numerically from the DNS models (Table \ref{table:velocities}).
The values for the P\'eclet numbers and other important parameters  required to simulate the particular initial, boundary, and derived conditions in the flow cell are summarized in Table \ref{table:velocities}. The value of $Pe_{\omega \omega}$ is slightly larger than unity,  indicating that convective transport and diffusive transport were both important in the inclusions. Although the intrinsic velocity in the inclusions ($||\langle v_{\omega} \rangle^{\omega}||$) is two orders of magnitude smaller than the intrinsic velocity in the matrix ($||\langle v_{\eta} \rangle^{\eta}||$), the $\omega$-region is characterized as {\it mobile} (values reported in Table \ref{table:velocities}).  However, as mentioned previously, on Fig.~2, this experiment plots near the boundary between the mobile-mobile and mobile-immobile regimes.  Thus, for this case, either model is likely to give acceptable results for simulating the breakthrough curve behavior; however, one might expect a mobile-mobile model to be somewhat more accurate (although at a higher computational cost).

\begin{table}[htp]
  \caption{Summary of macroscale system properties.}
  \begin{center}
    \begin{tabular}{c c c}
      \hline
      Parameter & Value & Units \\ \hline
    \\
	$||\langle {\bf v}_{\eta} \rangle^{\eta}||$ & $8.62 ~\times 10^{-6}$  & [m/s] \\
	$||\langle {\bf v}_{\omega} \rangle^{\omega}||$  &  $3.21 ~\times 10^{-8}$ & [m/s] \\
    $D^{**}_{L,\eta\eta}$ & $2.5\times 10^{-7}$ & [m$^2$/s] \\
    $D^{**}_{T,\eta\eta}$ & $2.5\times 10^{-8}$ & [m$^2$/s] \\
	$Pe_{\omega \omega}$ & $2.2$ & [ - ]\\
	$Pe_{\eta \omega}$ &  $14.8$  & [ - ] \\ 
    $Pe_{\eta \eta}$ &  $23925$  & [ - ] \\
    \\
    \hline
    \end{tabular}
  \end{center}
  \label{table:velocities}
\end{table}

Superimposed on Fig$.$~\ref{fig:btc_model_comparison} are the results from the fully-resolved DNS (model I), the reduced-domain DNS (model II), and the two upscaled models (models III and IV).  For this particular set of conditions, models III (two-equation, time-local) and IV (two-equation, quasi-steady) models produced results that were indistinguishable.  

Root-mean square (RMS) errors were computed for the difference between each of the four models and the experimental data for fluorescein as follows

\begin{equation}
\epsilon_{RMS} = \left( \sum_i [{\bar c}_{\eta,meas}(t_i) - {\bar c}_{\eta,model}(t_i)]^2  \right)^\frac{1}{2}
\end{equation}
For comparison purposes, we consider the ratio of RMS errors compared with the maximal-information model (model I) errors; for this metric, we refer to the RMS error associated with model I as $\epsilon_{0,RMS}$.

\begin{figure}
	\centering
  \includegraphics[scale=0.2]{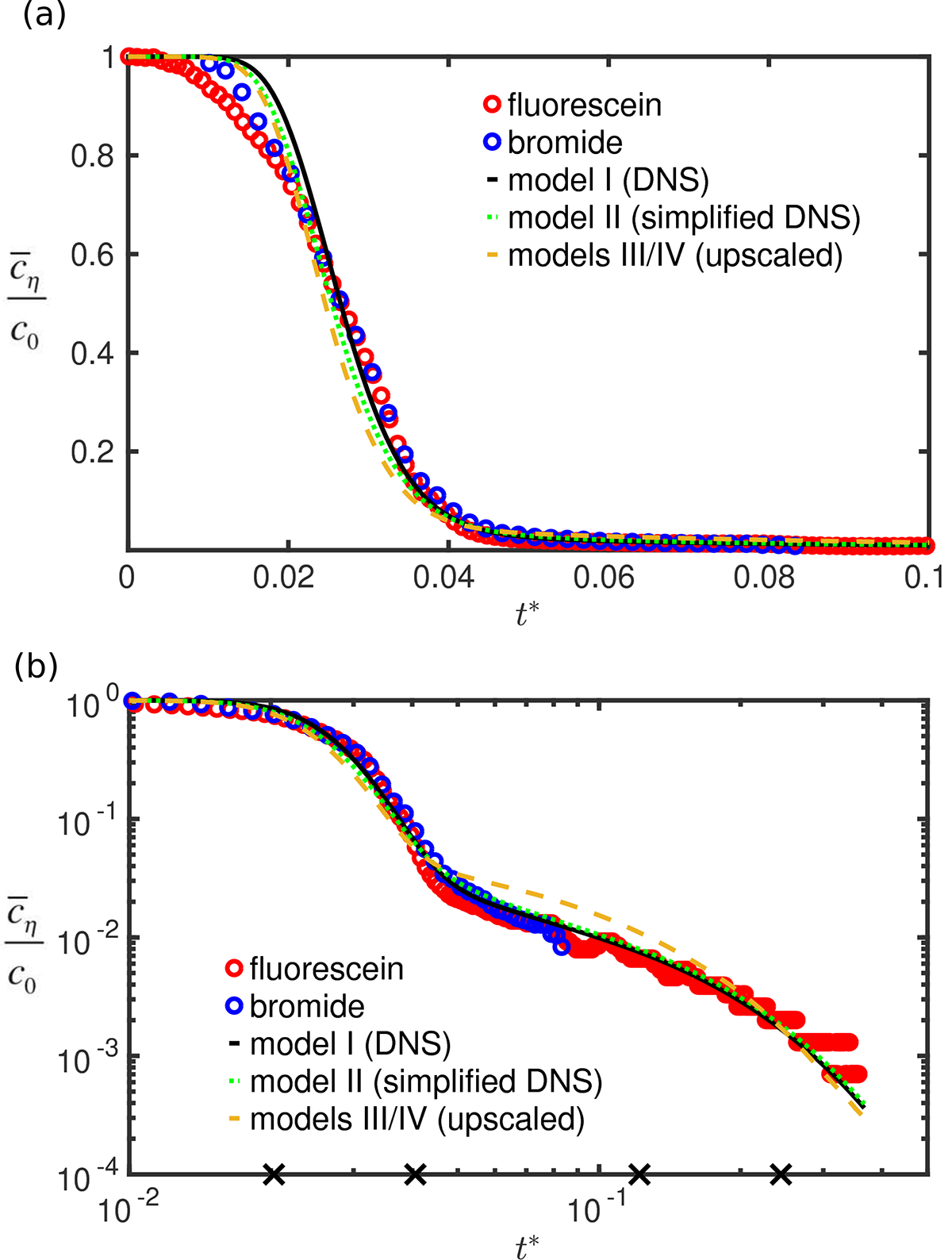}	
	\caption{Tracer breakthrough curves using (a) linear scaling, and (b) log-log scaling. 
	The $\times$ symbol on the time axis mark the times for which the concentrations in the system are visualized in Figure \ref{fig:conc_field_visualization}. 
	}
	\label{fig:btc_model_comparison}
\end{figure}

\begin{figure}
	\centering
  \includegraphics[scale=0.8]{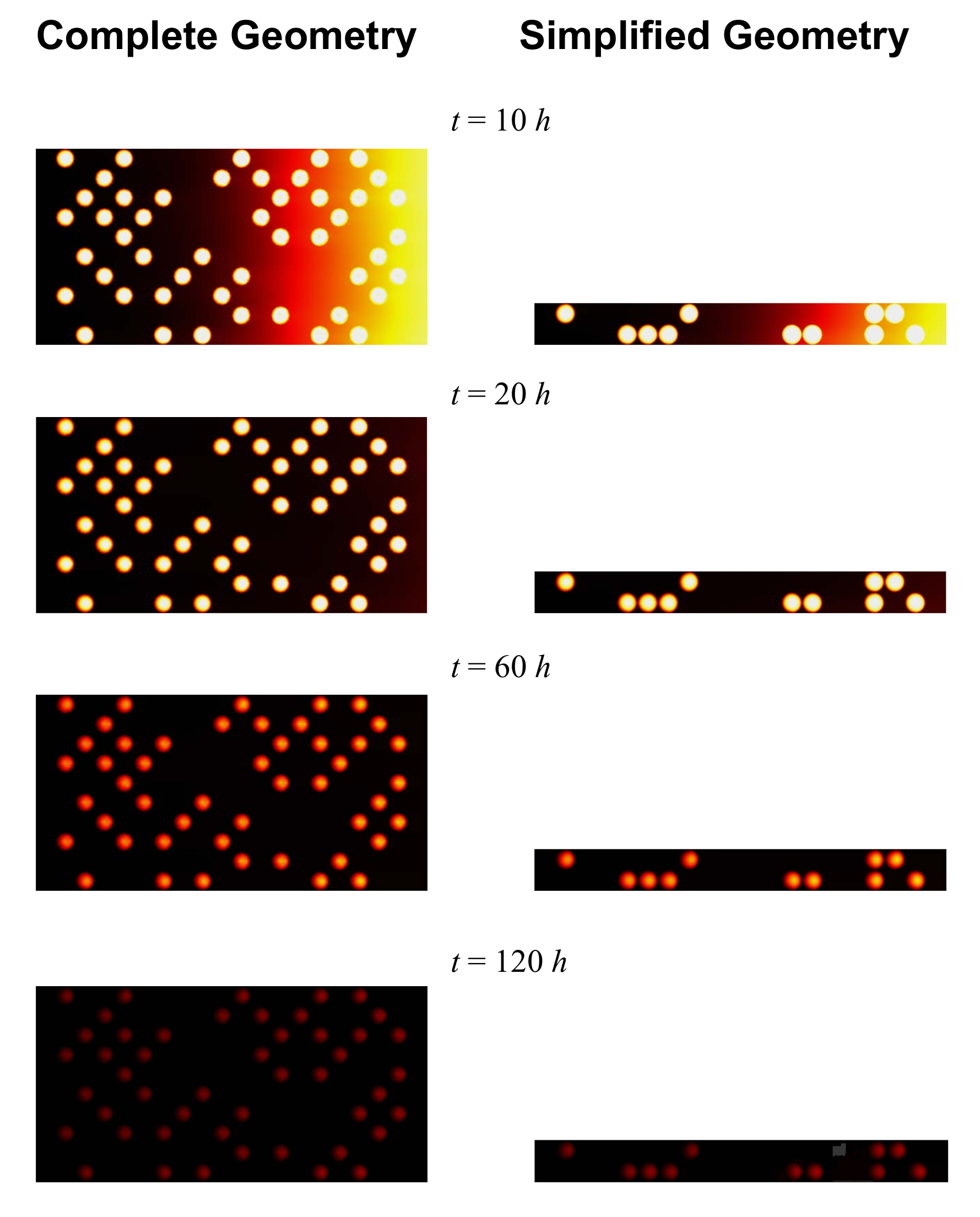}
	\caption{
	Visualization of DNS results of the fluorescein concentration field for both the fully-resolved and reduced-domain systems. The image planes and color scale are defined in Fig.~\ref{fig:flowcells}. Flow is from left to right.
	}
	\label{fig:conc_field_visualization}
\end{figure}

\section{Discussion} \label{sec:discussion}
\subsection{Model quality}

The results show reasonable agreement between the experiment and each of the the four models investigated. Each of the models explored predicted the trend of tracer breakthrough reasonably accurately.  It should be emphasized here that no numerical or parameter fitting was performed for any of the models;  all hydraulic parameters were either measured or computed for a unit cell model. The fully-resolved DNS and reduced-domain DNS produce almost identical breakthrough curves. This result also provides some validation that the measured parameters and geometrical representation for the flow cell are reasonably accurate, as evidenced by the very small RMS error associated with these simulations (Table \ref{tab:cpu}). An additional interesting feature of this modeling effort is that a purely Fickian-type constitutive equation was able to reproduce the system behavior accurately.  This is consistent with other recent work on inclusion-type laboratory systems (e.g., \citet{heidari_2014}).

Of all of the models, the fully-resolved DNS imposes the least reduction of complexity of the modeled problem. The model performance of reduced complexity domains (models II-IV) should therefore be compared to the fully-resolved DNS as well as the experimental data, since the early time behavior of the experimental data is not captured perfectly by any of the presented models (the most likely reason for this is experimental uncertainty rather than failure of any of the models; this is discussed below). The reduced-domain DNS captured the physics of the full system nearly as  accurately as the fully-resolved DNS. The only notable deviations to the fully resolved DNS are very small; again this is reflected in the very similar RMS error values presented in Table \ref{tab:cpu}. The increase in the RMS error was on the order of 5\% using the reduced-domain model; however, the domain size was only $1/10^{th}$ of the size of the fully-resolved DNS domain.

In comparison, the results obtained using the upscaled models show slightly larger deviations from the experiments than do the two DNS models. 
While the DNS correctly represents the time that the tailing begins (visible in Fig.~\ref{fig:btc_model_comparison}b), the volume averaged models predict a shift to the diffusive-dominated tailing regime slightly earlier. 
Subsequently, transport of solute from the inclusions to the matrix leads to higher concentrations at the outlet at the beginning of tailing, and a slight under-prediction of the solute concentration at the outlet at the end of the experiment.  Note that models III and IV are essentially indistinguishable for these particular conditions.  This observation is corroborated by noting that the mass transfer coefficient relaxes (i.e., where the two curves in Fig$.$~\ref{fig:alpha} are close enough to be considered indistinguishable) to its asymptotic value near about $t^* = 10^{-2}$.  In terms of non-normalized variables, complete relaxation  occurs after about 5 hours. This is a small fraction of the observed time, and thus transience in the value for the mass-transfer coefficient was not important for these conditions.

Another factor in the accuracy of models III and IV is the neglect of convection in the inclusions.  Analytical results such as those reported in \citet{haggerty_1995} and \citet{chastanet_2008} require that convection be neglected so that a simple analytical solution can be derived. Reference to Fig$.$~\ref{fig:pecletNumbers} shows that the experiment took place in the mobile-mobile region; thus, convection within the low-conductivity inclusions was not entirely negligible.  However, the plot in Fig$.$~\ref{fig:pecletNumbers} is intended to be primarily qualitative, especially near the regime boundaries; our data plots near such a boundary.  Thus, it is difficult to predict \textit{a priori} whether or not our experiments are significantly affected by convection.

For early times, each of the four models considered under-predicts the initial dispersion in concentration as compared to the experimental results. 
There are a number of possibilities here, but the most likely one is that the experimentally realized dispersion near the inlet of the flow cell is higher than it is in the remainder of the medium.  The most likely explanation is that there was a non-uniform distribution of flow into the flow cell. Although significant effort was taken to make the inlet as uniform as possible, inlet boundaries are notoriously difficult to control in porous media experiments.  Because we have no method to measure {\it a posteriori} the uniformity of the initial experimental injection, we have made no effort to account for this hypothesized early spreading in the models.  Regardless, the absolute difference for each of the models with the experimental data in this early part of the breakthrough curve is, at most, 5\% of the total concentration. \\

\subsection{Model complexity}

It is clear from the model results that using reduced-information models was effective for this experimental system, suggesting that the particular type and extent of heterogeneities in the system lends itself to upscaling. It is interesting to consider the question of \emph{how much} information reduction was realized by the various models employed.

One of the problems with computing information reduction in models is a fundamental one: what elements of a system can be considered to contribute to information for a model?  Because in this case, all four models are, ultimately, solved by numerical methods that require discretization of a set of partial differential equations, we might consider some of the tools that are used in measuring information content for algorithmic systems.  For example, one can consider computing the \emph{algorithmic complexity} of a particular discrete algorithm (note that in the literature, both \emph{information} and \emph{complexity} are used to indicate roughly the same concept).  The algorithmic complexity has a specific meaning in computing theory that can be well defined for particular problems \citep{zvonkin1970complexity,savage1976complexity,cook1983overview, cover2012elements}; however, it is neither easy to determine the algorithmic complexity for something as complicated as a finite element scheme \cite{farmaga2011evaluation}, nor does such a metric represent the only component of information that is of interest.  For example, the amount of memory required to solve a particular finite element problem is a relevant part of the complexity of a problem in applications \cite{farmaga2011evaluation} because memory limitations often define how well a particular problem can be resolved. However, algorithmic complexity does not always directly account for this facet of the problem \citep{cook1983overview}.  Algorithmic complexity also does not directly account for the amount of time it takes for a particular algorithm to run, although it is frequently proportional to it. Concepts such as Levin complexity \cite{ming1997introduction} address some of these problems, but it is still a tool that is best for more theoretical analyses.  

In many applications, it is more reasonable to develop a proxy for information content.  For iterative numerical methods, it has become common to use floating point operations or computational time or some filtered version of these metrics (e.g., \cite{karlsson2005complexity}) as a measure of complexity.  For our results, because we were using the same code and methods to solve all four models, we used a naive metric of computational time (CPU time) as a measure  of information content.  We did not make any attempts to account for overhead functions such as caching or inter-processor communications; however, a comparison of the data suggests that this metric scales linearly ($R^2=0.998$) with the number of degrees of freedom (DOF), suggesting that overhead functions were not significant for our particular problem.  Our results for processing times are summarized on Table~\ref{tab:cpu}.  Note that because models III and IV are essentially identical for our particular experimental conditions, they are not independently distinguished in the table.  In the remainder of this section, the terms \emph{complexity} and \emph{information} are used interchangeably; these terms are intended to be interpreted by the heuristic definitions discussed in the following rather than in the more formal definitions (associated with, for example, Shannon information or Kolmogorov complexity) that can be defined in the context of information theory.

\begin{table}[h!]
\caption{Computational demand for DNS and averaged models.
The value for the relative computational time (CPU \%) is calculated by normalizing to the time for Model I (fully-resolved DNS); the realtive RMS error is computed analogously.  Simulations were run on Intel(R) Xeon(R) CPU E5-2680 v3 processors. }
  \begin{center}
  {\footnotesize
  \begin{tabular}{ l r c c c c c c} 
  \hline 
  Model 	& DOF{\hspace{2mm}} 	& Time Steps 	&CPU Time & $f_t$ &$I_{CPU}$& $\epsilon_{RMS}$ & $\frac{\epsilon_{RMS}}{\epsilon_{0,RMS}}$\\
  $$		& $[-]$\hspace{3mm}	& $[-]$ & $[s]$ & $[-]$ & $[-]$ & $[-]$\\
  \hline 
  Model I 			& 1383630 	& \multicolumn{1}{r}{310} 	& \multicolumn{1}{r}{36248}  & \multicolumn{1}{r}{$1.0\times10^{0}~$}	& 0 & 0.0223  & 1.00\\
  Model II		& 152596 	& \multicolumn{1}{r}{297} 	& \multicolumn{1}{r}{2352} 	&  \multicolumn{1}{r}{$6.5\times10^{-2}$} & 1	& 0.0212 & 1.05\\
 Models III/IV& 194		& \multicolumn{1}{r}{73} 	& \multicolumn{1}{r}{4} 	   &  \multicolumn{1}{r}{$1.0\times10^{-4}$} 	& 4	& 0.0254 &  1.14\\
  \hline
  \end{tabular}
  }
  \label{tab:cpu}
  \end{center}
\end{table}

Information metrics are inherently difficult to compute, and it is conventional to examine changes in terms of powers of the base unit.  First, we define the relative time for computation compared to the model with the maximal use of information by

\begin{equation}
f_t = \frac{t_{m}}{t_0}
\end{equation}
where $t_0$ is the time for the maximum-information model to run, and $t_{m}$ is the time for any particular reduced-information model to run.  For our normalized metric, we can define the powers of 10 reduction in CPU time by

\begin{equation}
I_{CPU}=-log\left( \frac{t_{m}}{t_0}\right)
\end{equation}
This measure, then, provides the number of powers of 10 reduction in computation time.  We have computed these values, and listed them (rounding to the nearest log unit) in Table~\ref{tab:cpu}.  Because our information reduction is taken as a measure relative to the fully-resolved DNS, the fully resolved DNS model zero information reduction.  In other words, this model, when fully converged, represents the maximal use of information that we have regarding the experimental system, including the explicit details of its geometry, the boundary and initial conditions, and independent measures of the associated physical properties (Table~\ref{tab:system_properties}).  

In our efforts, employing models of reduced complexity is motivated by reducing the computational costs associated with systems of high variability of hydraulic parameters and complex geometries.  The results presented in Table~\ref{tab:cpu} provide some results about the relative computational costs of each of the four models.  First, we note that the fraction of computing time compared to the maximal-information model (model I) is substantially decreased in our hierarchy of reduced-complexity models.  Model II, which reduced the simulation domain by a factor of 10, correspondingly takes about a factor of 10 less time ($I_{cPU}=1$) to compute compared to model I. This comes at a cost in that the RMS error for model II is about 5\% higher than for model I; however, the reduction in computational burden may well make this trade off an acceptable one. 

The results for models III and IV show a substantially greater economy in computational costs.  The computational time for these models is on the order of 10,000 times less than for model I ($I_{cPU}=4$). Although the upscaled models show lower accuracy in respect to the experimental data than the fully-resolved and reduced-domain DNS (the RMS value for these models is about 14\% more than for model I), the results are still quite compelling when measured by the absolute RMS error, which is still quite small.

As a final comment, it is interesting to note that the question of reduction of complexity for models is incomplete without some notion of assigning utility to the various possible models. Without a notion of utility, it is impossible to make choices that weight the relative strengths of the models examined (in this case, the trade off is between computational size versus solution accuracy). This can be done by establishing a utility function that weights the various features of each model to provide an overall sense of its value (the review article by \cite{tartakovsky2013assessment} provides an excellent overview on this question).  For the cases explored here, the utility function would represent a trade-off between accuracy of the solution and the computational resources needed to obtain the solution.  Such issues become more acute when uncertainty is higher, and solutions may consist, for example, of large suites of Monte Carlo simulations in order to account for the uncertainty in the data.  

Even the understanding of what is meant by accuracy of the solution may have to be further quantified in a utility function.  For example, if one is primarily concerned about mean breakthrough time, then the arithmetic concentration, as plotted in Fig.~\ref{fig:btc_model_comparison}a might be relevant.  However, if one were more concerned about long-term behavior of small concentrations (this would be the case for some kinds of contaminants in the environment), then it may be the tailing of the breakthrough curves illustrated in Fig.~\ref{fig:btc_model_comparison}b that might be of more relevance.  The RMS values computed in this work were computed for the arithmetic concentrations, and would not be the most useful metrics of error if one wanted to emphasize the late-time tailing of the breakthrough curve.  Instead, comparisons of the various models might be best done on the log-transformed data, which would give increased weight to the late time data.  Thus, model simplification is one part of a larger analysis that must also consider how to place specific value (via a utility function or other method to assign value to outcomes) on the results that are obtained.

\section{Conclusion} \label{sec:conclusion}

A sequence of models with decreasing complexity was examined for modeling the breakthrough curves of a conservative tracer in a highly-heterogeneous, bimodal porous medium. 
The experimentally-observed late-time breakthrough was modeled using a two-region time non-local mass transfer model developed using the method of volume averaging that was cross-validated with direct numerical simulations at the Darcy scale. 
DNS with the full system geometry and a simplified geometry with identical volumetric fractions of two regions yielded very accurate results (as measured by RMS error) when compared with the experimental data for the entire time period examined experimentally. The DNS simulations included convective transport in the low-conductivity inclusions, and this proved to have a small, but measurable, impact on the accuracy of the results.

Breakthrough curves predicted by the two volume averaged models show only small differences from the two DNS results. 
These differences appear to arise almost entirely from the simplification in those models that neglects convection within the inclusions. Overall, the volume averaged models succeed in providing solutions require 4 orders of magnitude less computational time than the fully-resolved DNS, while only increasing the RMS error by a factor of 1.14.  For these particular cases, the absolute RMS error is already quite small for all of the models tested; thus a 14\% increase in the RMS error may not be of any practical significance.     

An interesting facet that the comparison of model complexity does raise is the question of how to decide what model is in, some sense, best.  Often in studies of this type, the stated goal is to develop reduced models that are computationally more efficient.  Considerations of model accuracy are usually relegated to not exceeding some (often arbitrary) standard of error.  In some sense, this does provide a {\it de facto} utility function for which a decision among models may be made.  However, in actual applications, one would have to attempt to assign value to the various costs involved for each of the models in order to make a selection.  In this example, the costs are primarily the interplay between computational time and model accuracy.  However, the specific objectives for the application of model predictions would potentially weight these two costs differently, and may even require  different metrics (e.g, a maximum concentration difference not to be exceeded at a control plane instead of the global RMS error) for measuring accuracy.  Although these questions are being increasingly recognized in hydrology \cite{tartakovsky2013assessment}, the development of concrete tools for assigning value to outcomes is clearly an important area for continued research.


\section{Nomenclature \label{nom}}
See Tables \ref{nom1} and \ref{nom2}.

\begin{table}
  \caption[Nomenclature.]{Nomenclature.}
  \label{nom1}
  \begin{center}
    \begin{tabular}{c c}
      \hline
      \hspace{5cm}\\
Roman symbols \\
      $a$					& Inclusion radius [$m$]\\
      $A_{\eta \omega}$			& Area of the interface between $\eta$- and $\omega$-region within an averaging volume V [$m^2$]\\
      $B$					& Kernel function defined by equation (\ref{eq:linear_alpha})\\
            ${\bf b}_\eta$					& closure variable for determining the effective dispersion tensor for the $\eta$-phase [$m$]\\
      $c_\eta$					&	Darcy-scale scalar concentration field, $\eta$-phase [$kg/m^3$]\\
            $c_\omega$					&	Darcy-scale scalar concentration field, $\omega$-phase [$kg/m^3$]\\
      $\left<c_\eta\right>^{\omega}$		&	$\omega$-region averaged  concentration field for use in models III and IV [$kg/m^{3}$]\\
      $\left<c_\omega\right>^{\eta}$		&	$\eta$-region averaged  concentration field for use in models III and IV [$kg/m^{3}$]\\
            $\bar{c}_\eta$					&	Flux-averaged (over flow cell effluent plane) concentration field, $\eta$-phase [$kg/m^3$]\\
      ${D_m}$	&	molecular diffusion coefficient for fluorescein [$m^2/s$]\\
      ${\textbf{\sffamily \bfseries D}}_{\mathit{\eta}}$	&	Darcy-scale effective diffusivity tensor, coarse medium [$m^2/s$]\\
         ${\textbf{\sffamily \bfseries D}}_{\mathit{\omega}}$	&	Darcy-scale effective diffusivity tensor, fine medium [$m^2/s$]\\
       ${\textbf{\sffamily \bfseries D}^*}_{\mathit{\eta}}$	&	Darcy-scale effective total dispersion tensor, coarse medium [$m^2/s$]\\
    ${\textbf{\sffamily \bfseries D}^*}_{\mathit{\omega}}$	&	Darcy-scale effective total dispersion tensor, fine medium [$m^2/s$]\\
      ${\textbf{\sffamily \bfseries D}}^{**}_{\eta \eta}$	&	Upscaled effective total dispersion tensor for the $\eta$-region [$m^2/s$]\\
            ${D}^{**}_{L,\eta \eta}$	&	Lateral upscaled total dispersion tensor component (the 1-1 component) [$m^2/s$]\\
                        ${D}^{**}_{T,\eta \eta}$	&	Transverse upscaled total dispersion tensor component (the 2-2 component) [$m^2/s$]\\
   $f_t$			&	Fraction of time for a model to run to completion compared to the maximal-information model [$-$]\\
      ${\textbf{\sffamily \bfseries I}}$	&	Second-order identity tensor [$-$]\\
      ${I_{CPU}}$	&	Number of (base 10) log units of difference in $f_t$ among models; \\
     $~$ &used as a heuristic measure of model complexity [$-$]\\
      ${\textbf{\sffamily \bfseries K}}_{\mathit{\eta}}$	&	Effective hydraulic conductivity tensor, coarse medium [$m/s$]\\
         ${\textbf{\sffamily \bfseries K}}_{\mathit{\omega}}$	&	Effective hydraulic conductivity tensor, fine medium [$m/s$]\\
      $K_{\eta}$ & Isotropic hydraulic conductivity of the $\eta$-region [$m/s$] \\
      $K_{\omega}$ & Isotropic hydraulic conductivity of the $\omega$-region [$m/s$] \\
            $L_m$					& Length of the porous material in the flow cell  [$m$]\\
      $L$					& Length of the flow cell (including inlet structures) [$m$]\\
            ${\bf n}_{\eta\omega}$					& normal vector for the $\eta-\omega$ interface, pointing from the $\eta$ phase toward the $\omega$  phase [$-$]\\
              ${\bf n}_{\eta e}$					& outward directed normal vector for the interface between the $\eta$ phase \\
      $~$ &and the exit of the flow cell [$-$]\\
            $Pe_{\omega\omega}$					& P\'eclet number for the fine material (inclusions) [$-$]\\
            $Pe_{\eta\eta}$					& P\'eclet number for the coarse material (matrix) [$-$]\\ 
                        $Pe_{\eta\omega}$					& Mixed P\'eclet number comparing convection in matrix to diffusion in inclusions [$-$]\\
    $t_0$			&	Time taken for the maximum-information model to run [$s$]\\
 $t_m$			&	Time taken for a reduced-information model to run [$s$]\\
      $\textbf{v}_0$			&	Inlet boundary intrinsic velocity for the $\eta$-phase [$m/s$]\\
      $\textbf{v}_\eta$			&	Darcy-scale intrinsic velocity field, $\eta$-phase [$m/s$]\\
      $\textbf{v}_\omega$			&	Darcy-scale velocity field, $\omega$-phase [$m/s$]\\
            $\langle{v}_\eta\rangle^\eta$			&	$\eta$-region intrinsic averaged velocity field for use in models III and IV [$m/s$]\\
      $\textbf{v}_\omega$			&	Darcy-scale velocity field, $\omega$-phase [$m/s$]\\
      $||\langle v_{\eta} \rangle^{\eta}||$ & Magnitude of the intrinsic velocity in the $\eta$-region [$m/s$]\\
      $||\langle v_{\omega} \rangle^{\omega}||$ & Magnitude of the intrinsic velocity in the $\omega$-region [$m/s$]\\
      $V_{\eta}$			& Volume of $\eta$-region within an averaging volume V. [$m^3$]\\
      $V_{\omega}$			& Volume of $\omega$-region within an averaging volume V. [$m^3$]\\
      $W$				& Effective mass transfer function [$kg/m^3/s$]\\
      \hline
    \end{tabular}
  \end{center}
\end{table}

\begin{table}
  \begin{center}
     \caption{Nomenclature (Continued)}
       \label{nom2}
    \begin{tabular}{c c}
      \hline
      $\hspace{5cm}$ & $\hspace{11cm}$\\
Subscripts \\
      $\eta$ & Denotes the region associated with the high-conductivity matrix \\
      $\omega$ & Denotes the region associated with the low-conductivity inclusions\\
\\
Greek symbols \\
      $\alpha$				& Mass transfer coefficient [$1/s$]\\
      $\alpha_{L, \eta}$		& Longitudinal dispersivity in the $\eta$-region [m]\\
      $\alpha_{T, \eta}$		& Transverse dispersivity in the $\eta$-region [m]\\
      $\epsilon$  & Relative root mean square error computed for the grid convergence study \\
      $\epsilon_{RMS}$  & Root mean square error computed for the difference between \\
      $~$ & the normalized concentration data and models [$-$] \\
      $\varepsilon_{\omega}$		&	$\omega$-region porosity [$-$]\\
      $\varepsilon_{\eta}$		&	$\eta$-region porosity [$-$]\\
      $\varphi_{\omega}$		&	Total volume fraction of the $\omega$-region [$-$]\\
      $\varphi_{\eta}$			&	Total volume fraction of the $\eta$-region [$-$]\\
      $\rho_{\eta}$ & Density of the $\eta$-region [$g/cm^3$] \\			
      $\rho_{\omega}$ & Density of the $\omega$-region [$g/cm^3$] \\
      $\tau_{D}$ & Characteristic timescale of diffusion in the inclusions [$-$]\\
\\
Abbreviations \\
DNS  & Direct numerical simulation \\
DOF & Degrees of freedom\\
REV  & Representative elementary volume \\
RMS & Root mean square \\
      \hline
    \end{tabular}
  \end{center}
\end{table}

\section*{Acknowledgements and Data}
The experiments discussed in this study were originally conducted by Stephanie Harrington (Savannah River Remediation LLC) as part of doctoral work at OSU; we appreciate the careful work conducted as part of that thesis. BDW, RP, and SO were supported on NSF EAR 1521441. The data associated with this paper is publicly available on the following open source electronic archive:  \url{http://hdl.handle.net/}.



\section*{Appendix Convergence Analysis}\label{sec:convergence_analysis}

A convergence analysis of the model was performed to ensure sufficient accuracy of the numerical scheme.
Convergence for both time and space was quantified by computing the area integral of the resident concentration at breakthrough plane.  
Breakthrough curves were calculated for time steps ranging from $5\times 10^2$ seconds to $1\times 10^5$ seconds and various pre-defined and user-defined mesh parameters.
A root mean square (RMS) error was then computed to quantify convergence behavior for the time step and mesh size.  The RMS error was defined by

\begin{equation}
	\epsilon = \sqrt{
    \mathop { \sum^{t=n}_{t=0} \left( C_{\mathit{0}} - C_{\mathit{test}} \right)^2}
  }
\end{equation}
where $n$ is the number of time steps in the simulation, $C_{0}$ is the concentration data of the breakthrough curve calculated with the smallest mesh and time step computed, and $C_{test}$ is the concentration data of the breakthrough curve for the mesh and time step in which convergence is being evaluated. 
The size of the RMS and the run time required to compute results for the model were the basis for choosing a sufficiently small mesh size and time step.  The system was assumed to be converged when reducing the mesh size or time step did not significantly lower the RMS value.
Variations of the time step had a small influence on the accuracy of the solution (Figure \ref{fig:convergence_analysis}a), and a time step of $1\times 10^4$ seconds was subsequently chosen. 
Convergence was defined by achieving an RMS value less than $\epsilon = 0.0025$; this value was met  for mesh edge lengths in the matrix around 0.02~m (Figure \ref{fig:convergence_analysis}b).  A maximum mesh edge length of 0.0212~m was used for the direct numerical simulations presented in the results (Section \ref{sec:results}). 

\begin{figure}
\centering
a) \includegraphics[width = .45\textwidth]{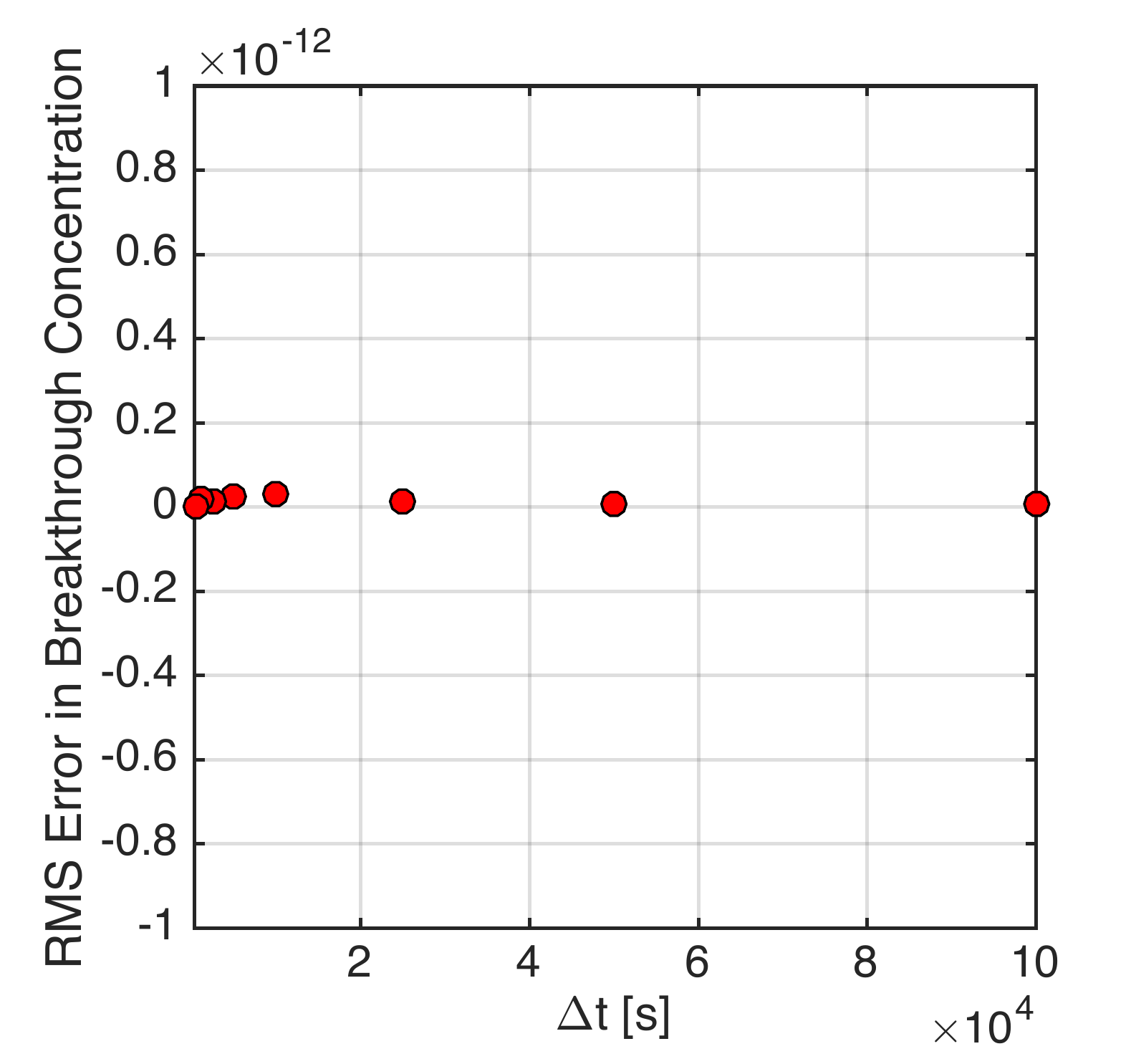} \quad
b) \includegraphics[width = .45\textwidth]{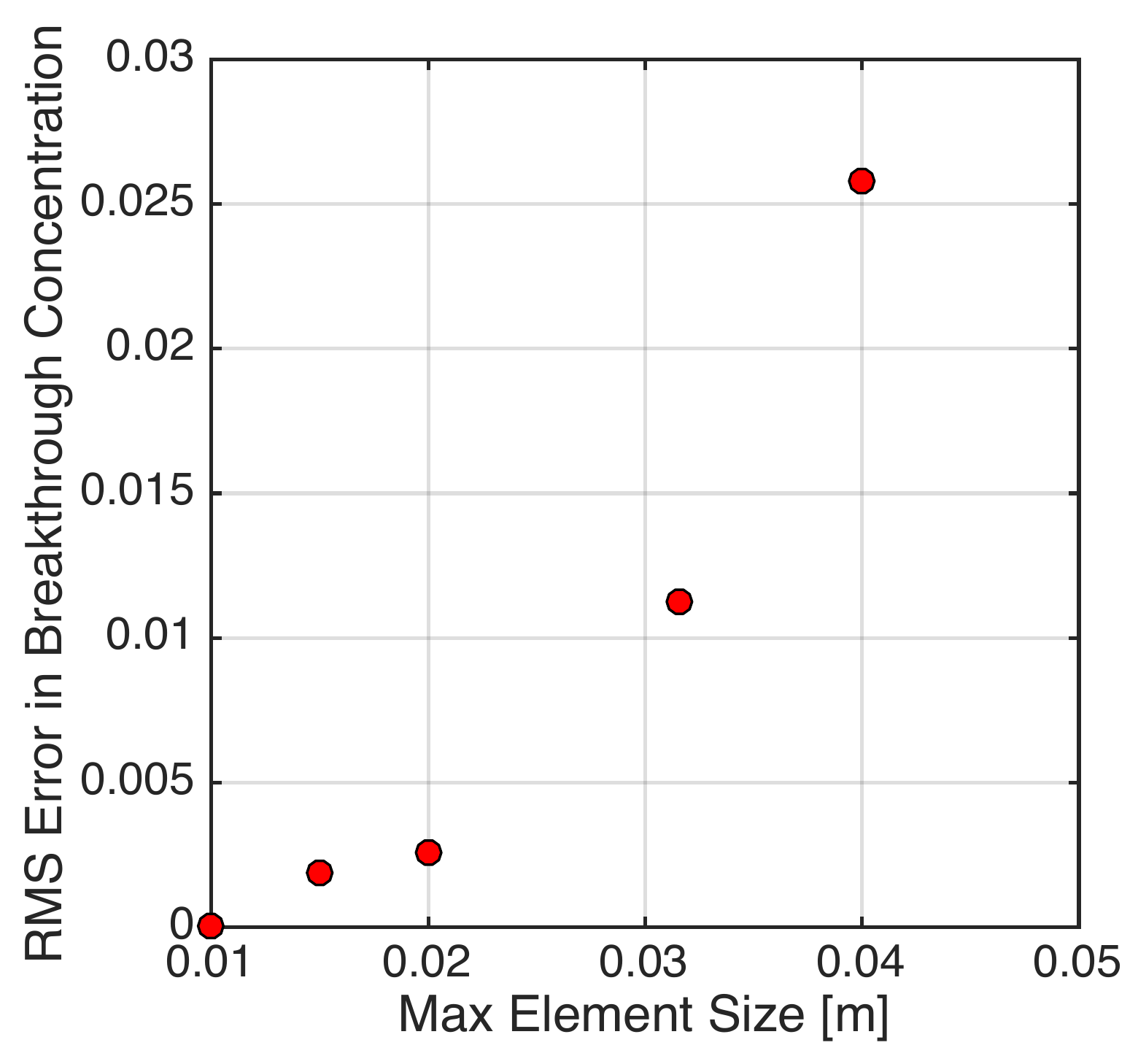}	
\caption{Relative RMS error in breakthrough concentrations for; a) the time step; and b) maximum mesh edge length. 
Convergence analysis for the time step was performed on the simplified system and convergence analysis for the mesh size was performed on the full system. 
}
\label{fig:convergence_analysis}
\end{figure}
\newpage
\bibliographystyle{elsarticle-harv.bst}      
\bibliography{bibliography}

\end{document}